\documentclass[11pt]{article}

\usepackage{amsfonts}
\usepackage{amsmath}
\usepackage{amsthm}
\usepackage{amssymb}
\usepackage[onehalfspacing]{setspace}
\usepackage{graphicx}
\usepackage{rotating}
\usepackage{pdflscape}
\usepackage{authblk}
\usepackage{color}
\usepackage[round]{natbib}
\usepackage{hyperref}
\usepackage{caption}
\usepackage{subcaption}
\usepackage{algorithm2e}
\usepackage[dvipsnames]{xcolor}
\usepackage{multirow} 

\usepackage{colortbl}
\colorlet{lg}{gray!20}
\newcommand{\sh}{\cellcolor{lg}}

\usepackage{tikz} 
\usetikzlibrary{shapes,arrows,shadows}

\allowdisplaybreaks[1]
\usepackage[flushleft]{threeparttable}

\definecolor{dkgreen}{rgb}{0,0.4,0}
\definecolor{gray}{rgb}{0.5,0.5,0.5}
\definecolor{mauve}{rgb}{0.58,0,0.82}

\begin{document}
\pgfdeclarelayer{background}
\pgfdeclarelayer{foreground}
\pgfsetlayers{background,main,foreground}
\tikzstyle{sensor}=[circle,draw=black, fill=LimeGreen,minimum height=1.5em,
line width=0.5mm]
\tikzstyle{ann} = [above, text width=15em, text centered]
\tikzstyle{wa} = [sensor,  fill=red,minimum height=1.5em]

\title{Structured factor copulas for modeling the systemic risk\\ of European and United States banks }
\author{
Hoang Nguyen$^{(a)}$, Audron\.e Virbickait\.e$^{(b)}$, M. Concepci\'on Aus\'in$^{(c)}$,\\ 
and Pedro Galeano$^{(c)}$\\
\begin{small}
$^{(a)}$Department of Management and Engineering, Linköping University, Sweden\\
$^{(b)}$Department of Quantitative Methods, CUNEF Universidad, Madrid, Spain\\
$^{(c)}$Department of Statistics and UC3M-Santander Big Data Institute, Universidad Carlos III de Madrid, Spain
\end{small}
}

\date{\today}
\maketitle

\begin{abstract}
In this paper, we employ Credit Default Swaps (CDS) to model the joint and conditional distress probabilities of banks in Europe and the U.S.\ using factor copulas. We propose multi-factor,  structured factor, and factor-vine models where the banks in the sample are clustered according to their geographic location. We find that within each region, the co-dependence between banks is best described using both, systematic and idiosyncratic, financial contagion channels. However, if we consider the banking system as a whole, then the systematic contagion channel prevails, meaning that the distress probabilities are driven by a latent global factor and region-specific factors. In all cases, the co-dependence structure of bank CDS spreads is highly correlated in the tail. The out-of-sample forecasts of several measures of systematic risk allow us to identify the periods of distress in the banking sector over the recent years including the COVID-19 pandemic, the interest rate hikes in 2022, and the banking crisis in 2023. 



\end{abstract}

\newpage


\section{Introduction} \label{sec:Introduction}

Gaining insight into the intricate patterns of co-dependence among financial institutions is crucial. According to \citet{OhPatton2017}, it became evident during the financial crisis of 2008 that the models employed to understand the relationships between financial assets were not sufficient. A more recent example of financial interconnectedness occurred in 2023 when three U.S. regional banks - Silicon Valley, Signature, and First Republic - as well as a European counterpart, Credit Suisse, collapsed. Some authors argue that the implosion of the Silicon Valley bank has triggered financial instability in European markets, a phenomenon referred to as financial contagion \citep{Akhtaruzzaman2023}. Financial contagion refers to the cascading effects of some initial shock to a small portion of a financial system can have on the entire system \citep{Dornbusch2000,Forbes2002,Constancio2012}. Modeling and forecasting contagion is of major importance: for instance, given one bank has defaulted, what is the probability that the others will default? This issue has received a great deal of attention, as evidenced by \cite{Acemoglu2015, Dungey2015, Martins2023}.

Modeling banking contagion is not a simple task, since it can arise from several sources, as widely documented by the finance-related literature \citep{Ballester2016,Kleinow2016,Agosto2020}. Most of the articles differentiate between two channels of contagion: systematic and idiosyncratic, also characterized in \citet{Agosto2020} as global and local, respectively. The systematic or global channel is associated with certain global (i.e., inter-country) factors, as seen in the 2008 financial crisis, while the idiosyncratic or local channel is related to some bank-specific (i.e., inter-institutional) factors, as witnessed in 2023. However, the question persists as to whether banking contagion results from shared underlying factors, indicating systematic contagion, or if it transmits from one bank to another, demonstrating an idiosyncratic nature.

The first contribution of this article is a novel approach to capture complex co-dependence patterns between financial institutions, enabling the identification of both sources of financial contagion, systematic and idiosyncratic (global and local, in other words). Figure \ref{fig:diagram} presents a simplified diagram of the two contagion transmission sources: the panel on the left illustrates the case of systematic contagion caused by some common factor $F$ that affects all banks $B_i$ in the system, meanwhile the panel on the right shows the case of idiosyncratic contagion, which is caused by a single bank $B_1$. In practice, it is highly probable that both channels operate simultaneously, with the intensity of each source fluctuating over time. Additionally, the presence of multiple common factors, an ambiguous factor structure, and uncertainty regarding the quantity and nature of connections between the nodes are additional considerations. In other words, we are facing a complex co-dependence structure, that is high-dimensional, (potentially) non-linear, and non-Gaussian. Therefore, in this article we propose to model the complex co-dependence structure between financial institutions using the factor copulas of \citet{Krupskii2013,Krupskii2015} and the factor-vine copulas of \citet{Brechmann2014} and \cite{Nguyen2019b}. In these models, the dependence structure is defined through latent variables using bivariate copulas. More precisely, we employ five model designs, starting with one- and two-factor copulas \citep{Krupskii2013}, that permit the units (banks) in the system to be affected by one or several latent factors. Then, we also consider alternative factor structures that do not suffer from the curse of dimensionality, namely bi-factor and nested-factor models \citep{Krupskii2015}. Nonetheless, the aforementioned setups solely accommodate unit-factor interaction, i.e.\ systematic contagion. To relax this assumption and enable inter-institutional interaction, we adopt the factor-vine copula model \citep{Brechmann2014,Nguyen2019b} that effectively captures both forms of financial contagion, systematic and idiosyncratic.

\begin{figure}
    \begin{center}
    \begin{minipage}{.4\textwidth}
\begin{tikzpicture}
    \node (wa) [wa]  {\scriptsize $F$};
    \path (wa.south)+(-2,-1) node (m1f)[sensor] {\scriptsize $B_1$};
    \path (wa.south)+(0,-1) node (m2f)[sensor] {\scriptsize $B_2$}; 
    \path (wa.south)+(2,-1) node (m3f)[sensor] {\scriptsize $B_3$}; 
    \path [draw, <-,very thick] (m1f.north) -- node [above] {} (wa.250);
    \path [draw, <-,very thick] (m2f.north) -- node [above] {} (wa.270);
    \path [draw, <-,very thick] (m3f.north) -- node [above] {} (wa.290);
\end{tikzpicture}
\end{minipage}
\begin{minipage}{.4\textwidth}
\begin{tikzpicture}
    \node (sensor) [sensor] (b1) at (0,0) {\scriptsize $B_1$};
     \node (sensor) [sensor] (b2) at (2,0) {\scriptsize $B_2$};
     \node (sensor) [sensor] (b3) at (4,0) {\scriptsize $B_3$};

     \draw[->, very thick] (b1) to[bend left] node[left] {} (b2) ;
     \draw[->, very thick] (b1) to[bend right] node[left] {} (b3) ;
\end{tikzpicture}
\end{minipage}
\end{center}
    \caption{Systematic and idiosyncratic contagion.}
    \label{fig:diagram}
        \caption*{{\scriptsize The plot on the left shows the systematic contagion caused by a common factor $F$ that affects all banks $B_i$ (inter-country or global), while the plot on the right shows the idiosyncratic contagion caused by a single bank $B_1$ (inter-institutional or local). }}
\end{figure}

The second contribution of this article is the use of the Variational Bayes (VB) estimation method for the proposed factor copulas. Variational Inference is a computational technique used in Bayesian statistics and machine learning to estimate complex probability distributions that allows for fast inference and prediction even with very large datasets, see \citet{Blei2017} for a review. In particular, we follow \citet{Kucukelbir2017} to approximate the posterior distribution of the copula model parameters. A very relevant feature of our proposed inference procedure is that it can also facilitate an automated model selection process for determining the type of bivariate relationships between the variables, something that cannot be easily done using e.g.\ maximum likelihood, for which these relationships must be established in advance \citep{Krupskii2013,Krupskii2015}.

The third contribution of this article is an empirical application using the most recent CDS (Credit Default Swap) data of European and U.S.\ banks. A CDS is a financial contract in which a buyer of corporate or sovereign debt reduces possible loss arising from the default of the issuer of such debt. 
The buyer pays the CDS premium regularly over the life of the CDS contract in exchange for protection against the default of a specific borrower.
The interest in modeling and forecasting CDS spreads stems from their efficacy as indicators of market perceptions on the likelihood of default for the reference entity, be it a company or a nation \citep{Creal2015, Ye2022}. As noted in  \citet{OhPatton2018}, the main factor responsible for CDS spread fluctuations is changes in the probability of default. Moreover, \citet{Ye2022} found that the significance of the CDS spread variable in predicting default probabilities is invariant to various firm types and market conditions. Therefore, modeling the joint distribution of the CDS spreads enables to model and forecast the joint distribution of default probabilities. Indeed, CDS spreads have been widely employed to measure and model financial and sovereign contagion, see for example \citet{Beirne2013, Alter2014, Chen2020, Agosto2020}. Furthermore, as demonstrated by \citet{Alter2012,DeBruyckere2013,Alter2014}, there exist risk spillovers between banks and sovereigns. Therefore, successful forecasting of default probabilities of banks can help regulators and policymakers to act accordingly in order to reduce the possible risk spillover toward sovereigns.

In the empirical illustration, we employ the daily CDS spread data for eight European and seven U.S.\ banks for a period of more than 16 years. The chosen period encompasses the 2008 global financial crisis, the 2009-2014 European sovereign debt crisis, the COVID-19 pandemic of 2020, the 2022 interest rate spikes, and the collapse of several financial institutions in 2023. To our knowledge, this is the longest (time-wise) CDS series considered in the default probability modeling context in the banking sector. Our findings indicate that the financial contagion between Europe and U.S.\ can be attributed to systematic channels only, where there is a single global factor affecting both markets and two regional factors affecting each market individually. Moreover, if we consider European and U.S.\ markets individually, the financial contagion sources change. Regional financial contagion is transmitted not only from some latent common factor (systematic) but also from one bank to another (idiosyncratic). By using the best-fitting bi-factor and factor-vine copula models we obtain 20-step-ahead forecasts for various measures of systematic risk. We show that our proposed models can accurately capture the distress periods actually observed in historical CDS spread series. 

The rest of the paper is organized as follows. Section \ref{sec:Related_literature} contains a brief review of the related literature on the use of copula in modeling financial interconnectedness. Section \ref{sec:Model_specification} introduces the marginal model for bank default risk and the factor copulas used for modeling the joint distress probabilities. We present the Variational Bayesian inference algorithm in Section \ref{section:bayesianinference}. Section \ref{sec:empirical} contains the application of the proposed factor copula for modeling the joint and conditional distress probabilities of 15 European and U.S.\ banks. Finally, conclusions are drawn in Section \ref{sec:conclusion}.

\section{Related literature}  
\label{sec:Related_literature}

\citet{Rodriguez2007} was one of the first authors to study financial contagion between countries/regions using copulas. Using East Asian and Latin American stock indices he found evidence of changing dependence during the periods of turmoil and increased tail dependence in some markets but not the others. \citet{Aloui2011} and \citet{Kenourgios2011} study financial contagion using copulas by analyzing the co-dependence between the developed markets, such as the U.S.\ and U.K., and emerging markets, in particular, the BRICs. \cite{Arakelian2012} rely on copulas to capture instances of financial contagion by employing a change-point approach. The authors propose to test the existence of financial contagion by examining whether the number of breaks in copula parameters and individual marginal volatilities is different from zero. This approach provides evidence of financial contagion in equity markets and between sovereign entities during both the Asian and Mexican crises.

Other authors, instead of focusing on country/region contagion, consider interconnectedness at the firm-level. \citet{Brechmann13b} study contagion in financial markets by using vine copulas. More precisely, the authors find evidence of asymmetric tail behavior and that U.S.\ banks have a stronger impact on the international financial markets as compared to Europe and Asia-Pacific using CDS spread data for $38$ large international banks and insurers for the period of 2006-2011. Similarly, \citet{Kleinow2016} investigate the drivers of systemic risk and contagion among European banks. The authors employ copulas to quantify the systemic risk contribution and systemic risk sensitivity by using the CDS spread data of 36 European financial institutions from 2005 to 2013. 
\citet{Cerrato2017} investigate joint credit risk in the U.K.\ banking sector by analyzing the weekly CDS spreads from 2007 to 2015. The study reveals a strong connection between the time-varying and asymmetric dependence structure of CDS log-differences and the joint probability of default. The authors employ a GAS-based dynamic asymmetric copula model. The results confirm that a significant part of the dependence structure of CDS spread log-differences is influenced by market factors. 

Closest to our proposed approach, \citet{OhPatton2018} employ a new class of one-factor dynamic copula to model the co-dependence of CDS spreads among 100  U.S.\ firms over the period 2006 to 2012. The authors find that while the probability of distress for individual firms has decreased since the height of the financial crisis, the current levels of joint probability of distress are significantly higher than those observed before the financial crisis. \citet{Krupskii2020} propose a time-varying factor copula model to capture the non-Gaussian co-dependence between the variables of interest. The authors apply their proposed model to the CDS spread data of 24 U.S.\ firms and provide estimates of the joint probability of distress, similar to the approach in \citet{Lucas2014} and \citet{OhPatton2018}.

\section{Model specification}
\label{sec:Model_specification}

This section describes in detail the models used to capture the interconnectedness of the default probabilities among financial institutions (banks). To this end, we first show how the CDS spreads are related to the default probabilities. We then use existing models to estimate the marginal distributions for the log-differences of the CDS spread data. Finally, we describe in detail the factor copulas used for modeling the joint behavior of the default probabilities. We define three classes of copula models: the multi-factor copulas of \citet{Krupskii2013}, which include the one and two-factor copulas; the structured factor copulas of \citet{Krupskii2015}, which include the bi-factor and nested copulas; and the novel factor-vine copulas of \citet{Nguyen2019b}, resulting in five distinct model architectures.

\subsection{Default probability and marginals}
\label{sec:marginal}

To assess default risk, one standard approach involves calculating the implied default probability by valuing the present value of both the premium payment and the protection payment. Following the simplified model in \citet{Lange2016}, let $S$ be the CDS spread, usually measured in basis points (bps), at time $t=0$ where the risk-free interest is $r$ and the risk neutral loss given default is $L$. The implied risk neutral default probability $p$ is the risk-neutral solution of premium payment and protection payment discounted for $\tau$ period of payment, i.e.
\begin{equation}
\frac{S}{1+r} + \frac{S}{(1+r)^2} + \ldots + \frac{S}{(1+r)^\tau} = \frac{pL}{1+r} + \frac{(1-p)pL}{(1+r)^2} + \ldots + \frac{(1-p)^{\tau-1}pL}{(1+r)^\tau}
\end{equation}
or equivalently,
\begin{equation} \label{eq:margdef}
\frac{ \left(1 - (1+r)^{-\tau} \right)S }{r} = pL \frac{1 - (1-p)^{\tau} (1+r)^{-\tau} }{r+p}.
\end{equation}
Here we assume that the default event only happens at the end of each payment period. The interest rate $r$ as well as the loss given default $L$ are constant. We choose the EURIBOR one year as the proxy of the risk-free interest rate and set $L=75\%$ as in \citet{Lucas2014}. Note that, when $r$, $p$ and $\tau$ are small enough, we could approximate $S \approx pL$.

The CDS spread series usually exhibit very high levels of autocorrelations, sometimes being very close to the non-stationary region, therefore, we consider the log-differences of daily CDS spreads as $s_{i,t}=\log\left(S_{i,t}\right)-\log\left(S_{i,t-1}\right)$ for each bank $i=1,\ldots,d$ and time unit $t=1,\ldots,T$. Additionally, we assume that CDS log-differences, $s_{i,t}$, follow a stationary, asymmetric autoregressive GJR-GARCH model with skewed and fat tail innovations. More precisely, we assume a AR($p$)-GJR-GARCH(1,1) model with standardized skew-$t$ errors given by:
\begin{equation}
\begin{aligned}
s_{i,t} & = \mu_i + \displaystyle\sum_{j=1}^p \phi_{j,i} s_{i,t-j} + a_{i,t}, \\
a_{i,t} & = \sigma_{i,t} \eta_{i,t}, \qquad \eta_{i,t}\sim \text{skew-$t$}(0,1,\xi_i,\nu_i),\\
\sigma_{i,t}^2 & = \omega_i + \left(\alpha_{i}+\gamma_i I_{a_{i,t-1}<0}\right) a_{i,t-1}^2  + \beta_{i} \sigma_{i,t-1}^2,
\end{aligned}
\end{equation}
where $\mu_i$ is a constant mean, $\phi_{j,i}$ are the autoregressive parameters, $\left(\omega_{i},\alpha_{i},\beta_{i},\gamma_i\right)$ are the GJR-GARCH parameters and $I_{a_{i,t-1}<0}$ indicates whether $a_{i,t-1}$ is negative, in which case it is 1, and when positive, in which case it is 0. To guarantee stationarity and positive variance, the root of the AR polynomial equation has to be less than one, $\left(\alpha_i+\gamma_i\right)>0$ and $0<\left(\alpha_i+\beta_i+\gamma_i/2\right)<1$. As a consequence, $a_{i,t}$ is a sequence of innovations or shocks, and  $\eta_{i,t}$ is a sequence of independent standardized innovations which is assumed to follow a standardized skew-$t$ distribution with skewness $\xi_i$ and degrees of freedom $\nu_i$, \citep{FernandezSteel1998}.

We note that any other alternative model can be specified for the marginals. \citet{Krupskii2013,Krupskii2015} employ an AR(1)-GARCH(1,1)-$t$ model. \citet{Creal2015} consider a stochastic volatility model with skew-$t$ errors, \citet{OhPatton2017} use an AR(1)-GJR-GARCH(1,1) model with a nonparametric error distribution, and \citet{OhPatton2018} employ an AR(5)-GJR-GARCH(1,1) model with skew-$t$ errors. These models can be different for each of the individual series. If the model is specified correctly, the probability integral transform results in uniformly distributed data, i.e., $F_{s_{i,t}}(s_{i,t})=u_{i,t}\sim U\left(0,1\right)$, where $F_{s_{i,t}}$ denotes the model distribution function for the CDS log-differences. Next, using the transformed variables $\mathcal{U}=\{u_{i,t}:i=1,\ldots d;t=1,\ldots T\}$, the co-dependence structure of the CDS spreads is modeled via different factor copula models.


\subsection{Co-dependence modeling via copulas}

The construction of flexible multivariate distributions using copulas started with the seminal work of \citet{Sklar59}. Since then, many studies have focused on utilizing these tools for various issues. For more information, please refer to \citet{ Joe2014}. A copula is a distribution on the multidimensional unit cube whose marginals are uniform. The use of copulas allows for the
separation of the modeling of the individual marginals from that of the co-dependence structure as follows. Consider a collection of random variables $Y_1,\hdots,Y_d$ with the corresponding marginal cumulative distribution functions $F_{Y_i}\left(y_i\right)=P\left[Y_i \leq y_i\right]$ for $i=1\hdots,d$ and a joint distribution function $H\left(y_1,\hdots,y_d\right)=P\left[Y_1\leq y_1,\hdots, Y_d\leq y_d\right]$. According to \citet{Sklar59}, there exists a copula $C$ such that
$H\left(y_1,\hdots,y_d\right)=C\left(u_1,\hdots,u_d\right)$, where $u_i=F_{Y_i}\left(y_i\right)$, for $i=1,\ldots,d$. It is straightforward that $u_i$ are realizations of a uniform random variable $U_{i} = F_i(Y_i)\sim U\left(0,1\right)$, for $i=1,\ldots,d$. As noted in \citet{OhPatton2017}, the copula decomposition property into margins, $F_i\left(\cdot\right)$, and dependence structure, $C\left(\cdot\right)$, has two important advantages over considering the joint distribution, $H\left(\cdot\right)$, directly. Firstly, estimation can be performed independently, greatly reducing the computational burden. Secondly, the choice of the marginal models is independent across series as well as from the co-dependence structure, as long as the models fit the data well. 

Unfortunately, finding adequate copula functions even in moderate dimensions may become complicated. One popular approach is the use of vine copulas where the dependence structure is defined as a product of bivariate copulas using tree representations, see \citet{Czado2022} for a review. However, the requirement of a significant number of parameters becomes an important problem when the dimension becomes high. Thus, truncated vine copulas and factor copulas are effective tools for capturing the dependence in high dimensional settings without requiring a large number of parameters. For example, \citet{Brechmann13b} assume that there is no dependence in higher tree level of vine copulas. Alternatively, \citet{Krupskii2013,Krupskii2015} propose factor copula models where observable variables link to unobservable variables via bivariate copulas. On the other hand, \citet{Creal2015} and \citet{OhPatton2017} extend the classical factor analysis by inverting the dependence structure from latent elliptical or skew-elliptical distributions to the constrained copula domain. In this paper, we concentrate on the former approach as it allows for a more flexible choice of copula functions. 

\subsubsection{Multi-factor copulas}
\label{sec:Factor_copulas} 

\citet{Krupskii2013} proposed a multi-factor copula model as truncated C-vines rooted at latent variables. Note that C-vines are vines whose trees have a main root node. They assume that the observables $U_1,\ldots,U_{d}$ are rooted at the latent variables $V_0,\ldots,V_{p-1}$ such that $U_i$, for $i =1,\ldots,p$, are independent given the latent factors $V_j$, for $j = 1,\ldots, p,$ which follow uniform distributions $V_j\sim U(0,1)$. Therefore, in each tree level of the factor copula model, a common factor stays at the root of the C-vine, and observable copula variables are linked to the common latent variable through bivariate copula functions.  
Table 1 in the Online Appendix shows some common bivariate copula functions as the building block for the one-factor and two-factor copula architectures which are illustrated in Figure \ref{fig:onetwofactorcopulas}. 
For example, let us denote each bivariate copula density as $c_{U_i,V_0}(u_i,v_0)$, each bivariate copula distribution function as $C_{U_i,V_0}(u_i,v_0)$ and the conditional distribution of $U_i$ given $V_0$ as $F_{U_i|V_0}(u_i|v_0)$ in the first tree level of the two-factor copula model.
The copula data in the second level of the two-factor copula are obtained as pseudo observables $u_{i|v_{0}}=F_{U_i|V_0}(u_i|v_0) \in (0,1)$, where $u_{i|v_{0}}$ are the realizations of a random variable $U_{i|V_{0}}$. The two-factor copula assumes that the variables in the second level $U_{i|V_{0}}$ are conditionally independent given the latent factor $V_1$. Hence, it is trivial to calculate the conditional copula density $p(u_1,\ldots, u_d|v_0,\ldots, v_p)$ and the integrated copula density $p(u_1,\ldots, u_d)$,  see details in Appendix \ref{app:fact}.

\begin{figure}[!htbp]
    \begin{center}
        \begin{minipage}{.45\textwidth}
\begin{tikzpicture}
\node(wa) [wa] (v0) at (0,0) {\scriptsize $V_0$};
\node (sensor) [sensor] (u1) at (-3,-1) {\scriptsize $U_1$};
\node (sensor) [sensor] (u2) at (-1.5,-1) {\scriptsize $U_2$};
\node (sensor) [sensor] (u3) at (-0,-1) {\scriptsize $U_3$};
\node (sensor) [sensor] (u4) at (1.5,-1) {\scriptsize $U_4$};
\node (sensor) [sensor] (u5) at (3,-1) {\scriptsize $U_5$};
\draw[ very thick] (v0) to  node[left] {} (u1) ;
\draw[ very thick] (v0) to  node[left] {} (u2) ;
\draw[ very thick] (v0) to node[left] {} (u3) ;
\draw[ very thick] (v0) to  node[left] {} (u4) ;
\draw[ very thick] (v0) to  node[right] {} (u5) ;
\end{tikzpicture}
\end{minipage}
\begin{minipage}{.45\textwidth}
\begin{tikzpicture}
\node(wa) [wa] (v0) at (0,0) {\scriptsize $V_0$};
\node(wa) [wa] (v1) at (-0.5,-1) {\scriptsize $V_1$};
\node (sensor) [sensor] (u1) at (-3,-2) {\scriptsize $U_1$};
\node (sensor) [sensor] (u2) at (-1.5,-2) {\scriptsize $U_2$};
\node (sensor) [sensor] (u3) at (-0,-2) {\scriptsize $U_3$};
\node (sensor) [sensor] (u4) at (1.5,-2) {\scriptsize $U_4$};
\node (sensor) [sensor] (u5) at (3,-2) {\scriptsize $U_5$};
\draw[ very thick] (v0) to[bend right]  node[left] {} (u1) ;
\draw[ very thick] (v0) to[bend right]  node[left] {} (u2) ;
\draw[ very thick] (v0) to[bend left=10]  node[left] {} (u3) ;
\draw[ very thick] (v0) to  node[left] {} (u4) ;
\draw[ very thick] (v0) to  node[right] {} (u5) ;
\draw[ very thick] (v1) to  node[left] {} (u1) ;
\draw[ very thick] (v1) to  node[left] {} (u2) ;
\draw[ very thick] (v1) to  node[left] {} (u3) ;
\draw[ very thick] (v1) to  node[left] {} (u4) ;
\draw[ very thick] (v1) to  node[right] {} (u5) ;
\end{tikzpicture}
\end{minipage}
\end{center}
    \caption{Multi-factor copulas.}
    \label{fig:onetwofactorcopulas}
    \caption*{{\scriptsize The plot on the left illustrates the one-factor copula, meanwhile the plot on the right illustrates the two-factor copula for $d=5$.}}
\end{figure}

\subsubsection{Structured factor copula}
\label{sec:bifcop}

To account for a more complex dependence structure and address the curse of dimensionality problem, \citet{Krupskii2015} propose an alternative approach by dividing $d$ variables into $G$ groups, so in total, there are $G+1$ latent variables  $V_0, V_{1}, \ldots, V_{G} \sim U(0,1)$. This gives rise to bi-factor and nested-factor copulas, see the illustration in Figure \ref{fig:binested}. The group division is arbitrary and is usually based on some external information, such as geographic location or firm sectors, for example. Alternatively, \citet{Oh2023} adapt a $k$-means clustering algorithm for the group division. 

In the bi-factor copula model, all $i_g$-th variables in group $g$ are conditionally independent given the latent variables, $V_{g}, V_0$, for $g = 1,\ldots, G,$ and $U_{i_g}$ does not depend on $g^{'}$, for any other $g' \neq g$, given $V_0$. Note that the bi-factor copula model is an extension of the two-factor copula, which is a bi-factor copula with $G = 1$. 
The nested-factor copula model assumes that $G$ latent variables, $V_1,\ldots,V_G$, are generated from a common latent variable $V_0$ and all observables $U_{i_g}$ in group $g$ are independent given the latent group factor, $ V_{g}$, for $g = 1, \ldots, G$. \citet{Krupskii2015} demonstrate the flexibility of the structured factor copula over the multi-factor copula models and derive the explicit form of copula density, see details in Appendix \ref{app:fact}.

\begin{figure}[!htbp]
    \begin{center}
        \begin{minipage}{.45\textwidth}
\begin{tikzpicture}
\node(wa) [wa] (v0) at (0,0) {\scriptsize $V_0$};
\node(wa) [wa] (v1) at (-1.5,-1) {\scriptsize $V_1$};
\node(wa) [wa] (v2) at (1.5,-1) {\scriptsize $V_2$};
\node (sensor) [sensor] (u1) at (-3,-2) {\scriptsize $U_1$};
\node (sensor) [sensor] (u2) at (-1.5,-2) {\scriptsize $U_2$};
\node (sensor) [sensor] (u3) at (-0,-2) {\scriptsize $U_3$};
\node (sensor) [sensor] (u4) at (1.5,-2) {\scriptsize $U_4$};
\node (sensor) [sensor] (u5) at (3,-2) {\scriptsize $U_5$};
\draw[ very thick] (v0) to[bend right] node[left] {} (u1) ;
\draw[ very thick] (v0) to  node[left] {} (u2) ;
\draw[ very thick] (v0) to node[left] {} (u3) ;
\draw[ very thick] (v0) to  node[left] {} (u4) ;
\draw[ very thick] (v0) to[bend left] node[right] {} (u5) ;
\draw[ very thick] (v1) to  node[left] {} (u1) ;
\draw[ very thick] (v1) to  node[left] {} (u2) ;
\draw[ very thick] (v1) to  node[left] {} (u3) ;
\draw[ very thick] (v2) to  node[left] {} (u4) ;
\draw[ very thick] (v2) to  node[left] {} (u5) ;
\end{tikzpicture}
\end{minipage}
\begin{minipage}{.45\textwidth}
\begin{tikzpicture}
\node(wa) [wa] (v0) at (0,0) {\scriptsize $V_0$};
\node(wa) [wa] (v1) at (-1.5,-1) {\scriptsize $V_1$};
\node(wa) [wa] (v2) at (1.5,-1) {\scriptsize $V_2$};
\node (sensor) [sensor] (u1) at (-3,-2) {\scriptsize $U_1$};
\node (sensor) [sensor] (u2) at (-1.5,-2) {\scriptsize $U_2$};
\node (sensor) [sensor] (u3) at (-0,-2) {\scriptsize $U_3$};
\node (sensor) [sensor] (u4) at (1.5,-2) {\scriptsize $U_4$};
\node (sensor) [sensor] (u5) at (3,-2) {\scriptsize $U_5$};

\draw[ very thick] (v0) to  node[left] {} (v1) ;
\draw[ very thick] (v0) to node[left] {} (v2) ;

\draw[ very thick] (v1) to  node[left] {} (u1) ;
\draw[ very thick] (v1) to  node[left] {} (u2) ;
\draw[ very thick] (v1) to  node[left] {} (u3) ;
\draw[ very thick] (v2) to  node[left] {} (u4) ;
\draw[ very thick] (v2) to  node[left] {} (u5) ;
\end{tikzpicture}
\end{minipage}
\end{center}
    \caption{Structured factor copulas.}
    \label{fig:binested}
    \caption*{{\scriptsize The plot on the left illustrates the bi-factor copula, and the plot on the right illustrates the nested-factor copula for $d=5$ and $G=2$.}}
\end{figure}

\subsubsection{Factor-vine copula} 

Finally, \citet{Nguyen2019b} extends the factor copula model proposed by \citet{Krupskii2013} to include a factor copula at the first tree level and truncated vine copulas at higher tree levels. The new proposed model aims to capture different behaviors at the tail of the joint distribution and
provides interpretable economic meaning. This idea is similar to \citet{Brechmann2014} who
construct Gaussian copulas using one-factor and Gaussian truncated vine, however, their
proposal ignores the issues of fat tails and asymmetric dependence among variables. 

Following the notations in \cite{Bedford2002}, \cite{Kurowicka06}, and \cite{Brechmann2012}, we specify the truncated factor-vine copula as a sequence of trees, and the nodes in each tree are linked by bivariate copula functions. Let $T_0, T_1, \ldots T_K$ be the trees in the truncated factor-vine copula model where $K \leq d-1$, and let $N_i$ and $E_i$ be the set of nodes and edges in tree $T_i$, for $i = 0, \ldots, K$, such that the following requirements are satisfied:
\begin{enumerate}
\item Tree $T_0$ has nodes $N_0 = \{0,1,\ldots,d \}$ and edges $E_0 = \{ \{0,1\}, \{0,2\}, \ldots, \{0,d\} \}$.
\item For $i = 1, \ldots, K$, the nodes in tree $T_i$ are the edges in tree  $T_{i-1}$, that is, $N_i = E_{i-1}$
\item If two edges in tree $T_i$ are joined by an edge in tree $T_{i+1}$, they must share a common node.
\end{enumerate}
The truncated factor-vine copula is a specification of truncated regular vine copulas (R-vine). Note that R-vines are vines only verifying requirements 2 and 3. Thus, truncated factor-vines are truncated R-vines where at tree $T_0$, there is a common root latent variable, $V_0$, that links to all observable variables by bivariate copula functions. Let the node set be $N = \{N_0, \ldots, N_K\}$ and let the edge set be $E = \{E_0, \ldots, E_K\}$. One associates each edge $e(j,k)$ in $E_i$ with a bivariate copula distribution function, $C_{j,k|D(e)}$, and a bivariate copula density, $c_{j,k|D(e)}$, where $j$ and $k$ are conditioned nodes and $D(e)$ is the conditioning set. The bivariate copulas in tree $T_0$ have an empty conditioning set.

Figure \ref{fig:fvcopulas} shows an example of a truncated factor-vine copula with two levels. Level 1 (on the left) contains a one-factor copula, and the remaining co-dependence among the variables, conditional on $V_0$, is captured by a regular vine copula on level 2 (on the right). 
Appendix \ref{app:fact} derives the conditional factor-vine copula density.

\begin{figure}
    \begin{center}
        \begin{minipage}{.45\textwidth}
\begin{tikzpicture}
\node(wa) [wa] (v0) at (0,0) {\scriptsize $V_0$};
\node (sensor) [sensor] (u1) at (-3,-1) {\scriptsize $U_1$};
\node (sensor) [sensor] (u2) at (-1.5,-1) {\scriptsize $U_2$};
\node (sensor) [sensor] (u3) at (-0,-1) {\scriptsize $U_3$};
\node (sensor) [sensor] (u4) at (1.5,-1) {\scriptsize $U_4$};
\node (sensor) [sensor] (u5) at (3,-1) {\scriptsize $U_5$};
\draw[ very thick] (v0) to  node[left] {} (u1) ;
\draw[ very thick] (v0) to  node[left] {} (u2) ;
\draw[ very thick] (v0) to node[left] {} (u3) ;
\draw[ very thick] (v0) to  node[left] {} (u4) ;
\draw[ very thick] (v0) to  node[right] {} (u5) ;
\end{tikzpicture}
\end{minipage}
\begin{minipage}{.45\textwidth}
\begin{tikzpicture}
\node(sensor) [sensor]  (u4) at (0,0) {\scriptsize $U_{4|V_0}$};
\node (sensor) [sensor] (u1) at (-3,-1) {\scriptsize $U_{1|V_0}$};
\node (sensor) [sensor] (u5) at (-1.5,-1) {\scriptsize $U_{5|V_0}$};
\node(sensor) [sensor]  (u2) at (1.5,-1) {\scriptsize $U_{2|V_0}$};
\node (sensor) [sensor] (u3) at (3,-1) {\scriptsize $U_{3|V_0}$};
\draw[ very thick] (u4) to  node[left] {} (u2) ;
\draw[ very thick] (u4) to [bend right] node[left] {} (u1) ;
\draw[ very thick] (u2) to  node[left] {} (u3) ;
\draw[ very thick] (u4) to  node[right] {} (u5) ;
\end{tikzpicture}
\end{minipage}
\end{center}
    \caption{Truncated factor-vine copula.}
    \label{fig:fvcopulas}
    \caption*{{\scriptsize The figure shows a truncated factor-vine copula with two levels for $d=5$. Level 1 on the left contains a one-factor copula while the remaining dependence of variables, conditional on $v_0$, is a regular vine copula on level 2 (on the right).}}
\end{figure}

\section{Bayesian inference}
\label{section:bayesianinference}

\citet{Krupskii2013,Krupskii2015} use maximum likelihood estimation for structured factor copula models. However, the parametric families of the bivariate copula links are chosen at the beginning which limits the flexibility of the model. In this section, we start with a factor copula with a known structure and connection links. We apply Variational Bayesian inference, similar to \citet{Nguyen2020}, to estimate the copula parameters. Then, we discuss an extension of an automated process to identify the dependence structure and to select the appropriate bivariate copula functions. 

\subsection{Prior and posterior distributions}

Assuming that we have a pre-specified factor copula structure together with bivariate linking copulas, the goal is to make inference on the latent variables $\mathbf v_{0:G, 1:T}$ and copula parameters $\boldsymbol \theta$ given the copula data $\mathcal{U}=\{u_{i,t}:i=1,\ldots d;t=1,\ldots T\}$.
For example, in the truncated two-level factor-vine copula model, let $\boldsymbol \Theta = \{ \mathbf  v_{0, 1:T}, \boldsymbol \theta \}$ be the set of parameters of interest, where $\mathbf  v_{0, 1:T}= \{v_{0,1},\ldots,v_{0,t}\}$ and $\boldsymbol\theta =\left\{\boldsymbol \theta_{01},\ldots, \boldsymbol \theta_{0d} \text{ and } \boldsymbol \theta_{jk},  \text{ for } e(j,k)\in E_1\right\}$. We assume a uniform distribution $U(0,1)$ for the latent variables, $v_{0,t}$, and an uninformative but proper prior distribution for any set of bivariate copula parameters in $\boldsymbol \theta$ so that the Kendall-$\tau$ is in the range of $[-0.9;0.9]$. See the details of prior distributions for copula parameters in Table 1 of the Online Appendix.

The prior density is a product of independent prior distributions of the model parameters, $\displaystyle \pi\left(\boldsymbol\Theta\right)=\prod_{t=1}^{T}\pi\left(v_{0,t}\right)\prod_{i=1}^{d}\pi\left(\boldsymbol\theta_{0i}\right)\prod_{e\{j,k\}\in E_1}\pi\left(\boldsymbol\theta_{jk}\right)$. Using the conditional copula density in Appendix \ref{app:fact}, the posterior distribution up to a normalizing constant of the truncated two-level factor-vine copula can be written as,
\begin{equation}
\begin{aligned}
\pi(\boldsymbol \Theta | \mathcal{U} ) &\propto  \prod_{t=1}^T p(u_{1,t}, \ldots, u_{d,t} | v_{0,t}, \boldsymbol  \theta) \pi(\boldsymbol \Theta) \\
&\propto  \prod_{t=1}^T \left[ \prod_{e\{j,k\} \in E_1} c_{U_{j}, U_k \mid V_0}\left(u_{j,t \mid v_{0,t} }, u_{k,t \mid v_{0,t} } ; \boldsymbol{\theta}_{jk} \right) \prod_{i=1}^{d} c_{U_{i}, V_0}\left(u_{i,t}, v_{0,t} ; \boldsymbol{\theta}_{0i}\right) \right]  \times \\
& \hspace{0.5cm} \prod_{t = 1}^T \pi(v_{0,t})  \prod_{i = 1}^d \pi(\boldsymbol \theta_{0i})  \prod_{e\{j,k\} \in E_1} \pi(\boldsymbol\theta_{jk}).
\end{aligned}
\end{equation}
The truncated factor-vine copula requires obtaining a new set of pseudo data, $u_{i,t|v_{0,t}} = F_{U_{i,t}|V_{0}}(u_{i,t}|v_{0,t})$, whenever we generate a new sample of any element in $\boldsymbol \Theta$ which makes the implementation of the Markov Chain Monte Carlo (MCMC) computationally expensive. Hence, in the next section, we employ an approximate approach to obtain samples from the posterior distribution, namely, Variational Bayes.

\subsection{Variational inference}
\label{sec:VB}

Variational inference is a computational technique used in Bayesian statistics and machine learning for approximating complex probability distributions. For the factor copula models, described above, we are interested in making inferences on the factor copula parameters $\boldsymbol \Theta$ based on the posterior distribution $\pi(\boldsymbol \Theta | \mathcal{U})$. Variational Bayes proposes a family of distributions $\mathcal{Q}$ and finds a proposal distribution $q(\cdot; \boldsymbol \lambda) \in \mathcal{Q}$, parameterized by a vector $\boldsymbol \lambda$, to minimize the Kullback-Leibler divergence between $q(\boldsymbol \Theta; \boldsymbol  \lambda)$ and $\pi(\boldsymbol \Theta | \mathcal{U})$. However, as $\pi(\boldsymbol \Theta | \mathcal{U})$ is defined up to a normalizing constant, VB instead maximizes the evidence lower bound ($\mathtt{ELBO}$):
\begin{equation*}
\begin{aligned}
\mathtt{ELBO}(\textcolor{dkgreen}{q}) &= \mathbb{E}_{\textcolor{dkgreen}{q}}[\mathtt{log} \textcolor{mauve}{p(\mathcal{U}, \boldsymbol \Theta)} ] - \mathbb{E}_{\textcolor{dkgreen}{q}}[\mathtt{log} \textcolor{dkgreen}{q(\boldsymbol \Theta; \boldsymbol  \lambda)}] \\
&= \mathtt{log} p(\mathcal{U}) - KL(\textcolor{dkgreen}{q(\boldsymbol \Theta; \boldsymbol  \lambda)} || \textcolor{mauve}{\pi(\boldsymbol \Theta | \mathcal{U})}) \leq \mathtt{log} p(\mathcal{U})
\end{aligned}
\end{equation*}
such that when $\textcolor{dkgreen}{q(\boldsymbol \Theta; \boldsymbol  \lambda)} = \textcolor{mauve}{\pi(\boldsymbol \Theta | \mathcal{U})}$, we obtain $\mathtt{ELBO} = \mathtt{log} p(\mathcal{U})$. 
Therefore, VB turns Bayesian inference into an optimization problem of finding $\boldsymbol  \lambda$ to maximize the $\mathtt{ELBO}$. 
 
\citet{Blei2017} review different approaches in choosing the variational distributions and optimization techniques. As the posterior distribution of the structured factor copula is complex, we follow \citet{Kucukelbir2017} and use a black box variational inference based on stochastic approximation of $\mathtt{ELBO}$. The general framework is to sample from the distribution $q(\boldsymbol \Theta; \boldsymbol  \lambda)$ and use these values to approximate the expectations in $\mathtt{ELBO}$ using Monte Carlo integration. 
In order to find the optimal $\mathtt{ELBO}$, we also sample from $q(\boldsymbol \Theta; \boldsymbol  \lambda)$ and take the average of the derivatives of $\mathtt{ELBO}$ with respect to the parameters and update the variational parameters according to the gradients. 
For example, in the truncated factor-vine copula presented in the previous section, the mean-field approximation assumes that $\mathcal{Q}$ is a Gaussian distribution and 
\begin{equation}
\begin{aligned}
q(\boldsymbol \Theta; \boldsymbol  \lambda) & = \prod_{t = 1}^T \pi(v_{0,t}; \boldsymbol  \lambda)  \prod_{i = 1}^d \pi(\boldsymbol\theta_{0i}; \boldsymbol  \lambda)  \prod_{e\{j,k\} \in E_1} \pi(\boldsymbol\theta_{jk}; \boldsymbol  \lambda),
\end{aligned}
\end{equation}
where $\boldsymbol \lambda$ contains the posterior mean and posterior variance of the Gaussian distribution. 
\citet{Nguyen2019b} presents multiple simulation studies regarding the trade-off between the computational speed and estimation accuracy using VB as compared to MCMC for truncated factor-vine copula models. The author shows that VB approximates quite well the posterior distribution of the latent variable and slightly underestimates the standard deviations of the bivariate copula parameters.

\subsection{Model selection}
Given a pre-specified factor copula structure, it is fast to estimate the model parameters using VB. However, the parametric families of the bivariate copula links are usually unknown due to the nature of the latent variables.  
For the structured factor copula model, \citet{Krupskii2015} use graphical tools and \citet{Fan2023} use a proxy estimation to define the bivariate copula links between variables. On the other hand, \citet{Nguyen2020} derive an automatic procedure to recover the hidden dependence structure. This procedure iterates between the estimation of the latent common factors and the copula model selection between the latent factors and the observables. For example, we first assign all Gaussian bivariate copulas to the structured factor copula and estimate the common latent factors using VB. Given the posterior medians of the latent factors, we consider other copula models and select the best family for the bivariate links between the observables and the latent factors using the Bayesian information criterion (BIC). We reassign the chosen copula families for the links and reestimate the model until the bivariate links remain unchanged or until a certain number of iterations is reached. \citet{Nguyen2020} show that this procedure increases the BIC of the factor copula model. Also, the accuracy rate of the correct links is above $70\%$ in high dimensional simulations with very similar log-likelihoods as compared to true simulated models. For the truncated factor-vine copula models, we can select the bivariate copulas jointly among levels or level-by-level, which is similar to \citet{Dissmann13}. We use a one-factor copula model to choose the bivariate links in the first level and apply the algorithm in \citet{Dissmann13} for higher levels given the pre-specified copulas in the first level. \citet{Nguyen2019b} shows that the accuracy rates are also high in all levels of the factor-vine copulas.

\section{Empirical illustration}
\label{sec:empirical}

\subsection{Data}
\label{sec:data}

In this section, we employ the proposed factor copulas to analyze the joint distress probability of European and U.S.\ banks. First, we compile a list of the largest European and U.S.\ banks based on total assets. We then select only the banks that have actively traded in CDS during the designated period between January 1, 2007 and May 1, 2023 (16 years and 4 months). Finally, we extract daily data for \textit{5-year maturity CDS of senior unsecured debt} for 8 European and 7 U.S.\ banks from Bloomberg.
The ticker symbols and full names are summarized in Table \ref{table:banks}. 


\begin{table}[!htbp] 
\centering
\caption{Bank data.} 
\label{table:banks}
\scalebox{0.8}{
\begin{threeparttable}
\begin{tabular}{ll|ll}
  \hline
  \multicolumn{2}{c|}{European banks} & \multicolumn{2}{c}{U.S.\ banks} \\ 
Ticker symbol & Full name & Ticker symbol & Full name \\
  \hline
BCS & Barclays & AXP & American Express Company  \\ 
  BNP & BNP Paribas & BAC & Bank of America Corporation  \\ 
  CS & Credit Suisse & C & Citigroup \\ 
  GLE & Societe Generale & GS & The Goldman Sachs Group  \\ 
  ISP & Intesa Sanpaolo & JPM & JPMorgan Chase  \\ 
  SAN & Santander & MS & Morgan Stanley  \\ 
  UBS & UBS & WFC & Wells Fargo \\ 
  UCG & UniCredit & &  \\
   \hline
\end{tabular}
\begin{tablenotes}
\item {\footnotesize\textbf{Notes:} The table reports main information for the European and U.S.\ banks: ticker symbol, full name and the main operating market.}
    \end{tablenotes}
      \end{threeparttable}  
}

\end{table}

\begin{figure}[!htbp]
\centering
\includegraphics[width=0.8\textwidth]{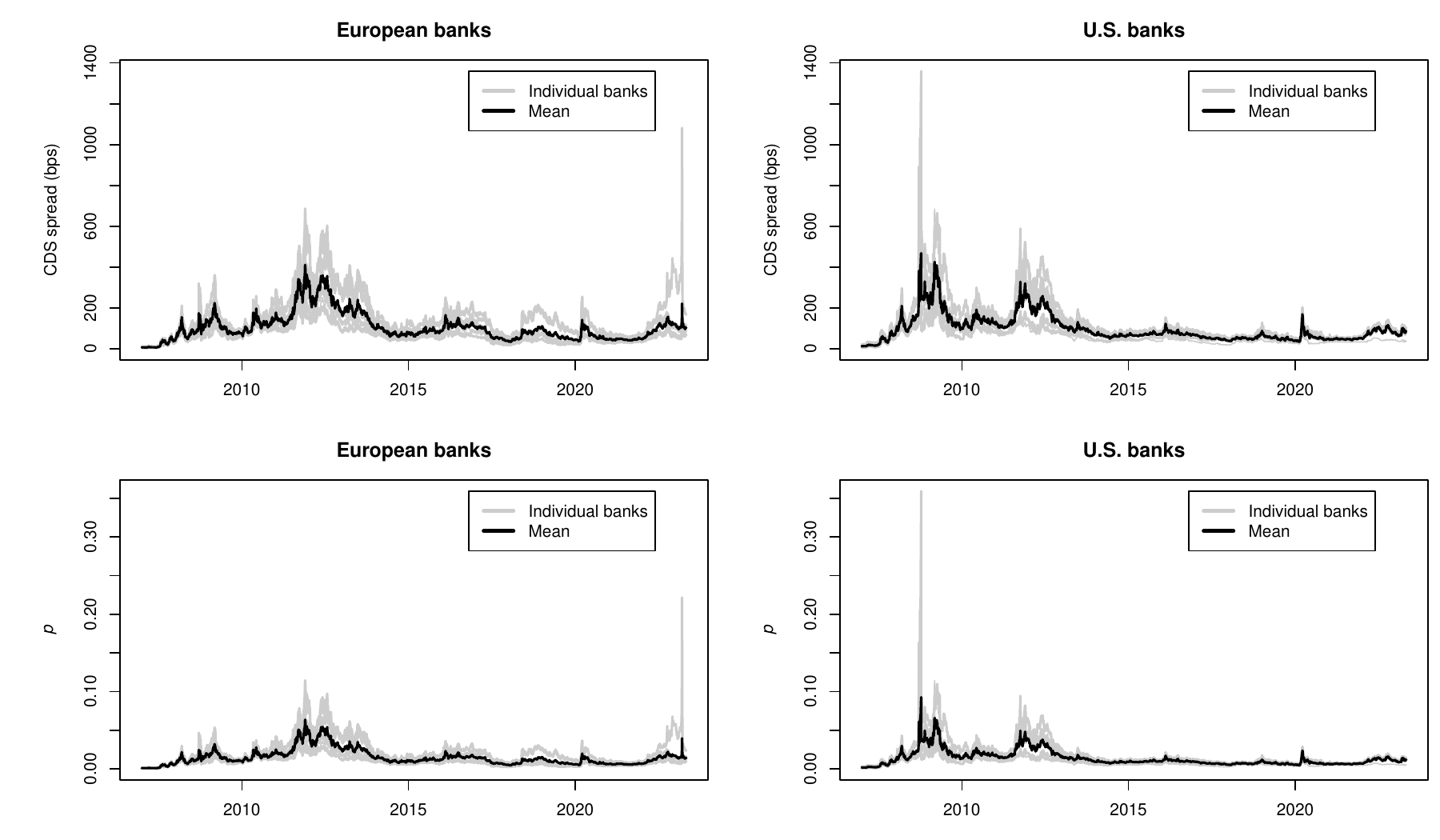}
\caption{CDS spreads and implied default probabilities by region.}
\label{fig:europe_us}
\caption*{{\scriptsize The figures show the evolution of the CDS spreads measured in basis points (top row) and the evolution of the implied default probabilities (bottom row) for January 2007 to May 2023. The European region contains 8 banks and the U.S.\ has 7 banks in total. The dynamics of each individual bank are in gray, and the average is in black.}}
\end{figure}

Before proceeding, we impute some missing values for some CDS spread series, see details in Online Appendix. After the treatment, we have a sample of $T = 4110$ observations. Figure \ref{fig:europe_us} draws the dynamics of the CDS spreads, measured in basis points, and the implied marginal default probabilities, obtained using Equation \ref{eq:margdef}. For both regions, various significant increases in CDS spreads indicate critical financial stress periods within the banking sector. The first increase is during the global financial crisis, where the U.S.\ financial sector suffered a greater impact as compared to the European banking sector. The second visible increase is during the turbulent developments in the European sovereign debt market during 2009-2014 \citep{Kleinow2016}, which affected both, Europe and U.S.\ banking sectors. As noted in the Financial Times, ``Credit default swaps on major U.S.\ banks (...) hit fresh highs for the year on Tuesday as problems in Greece intensified."\footnote{\url{https://www.ft.com/content/515508a2-9edc-11e1-9cc8-00144feabdc0}}. There is another notable increase at the year 2020 mark, which corresponds to the COVID-19 pandemic outbreak. Finally, the last peak in the CDS spreads is at the very end of the sample, which reflects the effect of the 2022 interest rate spikes on the banking sector and the early 2023 banking panic. The European region has a more pronounced increase because Credit Suisse bank is included in the sample, meanwhile, the CDS spread data for the three bankrupt U.S.\ banks is not available and, therefore, not reflected in the plot. Major descriptive statistics for the crude CDS spread data are reported in Online Appendix. The minimum values of CDS spreads vary between 5 and 20 bps, meanwhile, the maximum values go as high as 1360 bps for Morgan Stanley during the global financial crisis and 1082 bps for Credit Suisse in 2023. The bottom row of Figure \ref{fig:europe_us} draws the corresponding implied default probabilities for both regions. We can observe that the implied probabilities closely resemble the CDS spread dynamics. For both regions, the highest average implied default probabilities were reached during the worldwide financial crisis, and they have not reached those levels ever since (at least for the banks in the sample). 

\begin{figure}[!htbp]
\centering
\includegraphics[width=0.8\textwidth]{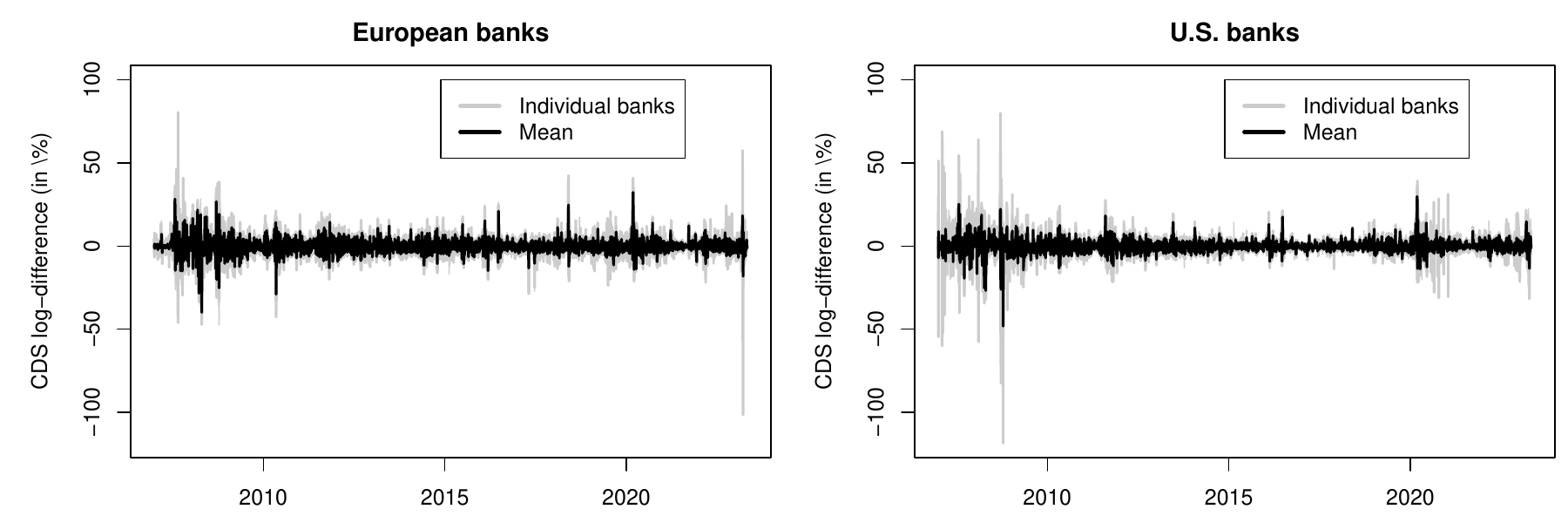}
\caption{CDS log-differences by region.}
\label{fig:lrets_europe_us}
\caption*{{\scriptsize The figures show the evolution of the CDS spread log-differences (in \%) from January 1st, 2007 to May 1st, 2023. European region contains 8 banks and the U.S.\ has 7 banks in total. The dynamics of each individual bank are in gray, and the average is in black.}}
\end{figure}

Next, the crude CDS spread data is converted into the log-differences $s_{i,t}$, as described in Section \ref{sec:Model_specification}. Table \ref{table:desc_lret} reports the summary statistics for the log-differences (in \%) of the CDS spreads of all 15 banks, and Figure \ref{fig:lrets_europe_us} plots the evolution of the log-differences. Most of the variables exhibit positive skewness and excess kurtosis, similar to the results in \citet{OhPatton2018}. Interestingly, the mean of log-differences is positive for all banks, with annualized averages between 7.5\% and 17.6\%. Positive CDS log-differences during this period mean that the CDS spreads tend to increase rather than decrease, on average.  In other words, the CDS spreads have increased on average for all banks during 2007-2023. Note that in some cases the log-differences are as low as negative 100\%. This does not pose any problems in the analysis, as the log transformation is used for mathematical convenience only. In the modeling and forecasting of systematic risk in Section \ref{sec:systrisk_results} we work with the CDS spread data and not the log-differences. 
 
\begin{table}[!htbp] 
\centering
\caption{Descriptive statistics for the log differences (in \%) of CDS spreads.} 
\label{table:desc_lret}
\scalebox{0.8}{
\begin{threeparttable}
\begin{tabular}{rrrrrrrrrrrr}
  \hline
 & mean & Q2 & sd & skew & kurt & min & max & 1stACF & ADF(10) & LB(10) & ARCH(10) \\ 
  \hline
BCS & 0.07 & -0.05 & 4.26 & -0.22 & 17.23 & -47.30 & 35.74 & 0.16 & 0.01 & 0.00 & 0.00 \\ 
  BNP & 0.06 & -0.04 & 4.53 & 0.17 & 11.03 & -31.51 & 34.86 & 0.12 & 0.01 & 0.00 & 0.00 \\ 
  CS & 0.07 & -0.02 & 4.45 & -2.35 & 85.99 & -101.62 & 57.53 & 0.18 & 0.01 & 0.00 & 0.00 \\ 
  GLE & 0.06 & -0.04 & 4.45 & 0.40 & 13.13 & -39.20 & 40.96 & 0.12 & 0.01 & 0.00 & 0.00 \\ 
  ISP & 0.07 & -0.03 & 5.01 & 1.26 & 31.68 & -46.07 & 80.52 & 0.06 & 0.01 & 0.00 & 0.00 \\ 
  SAN & 0.05 & -0.05 & 4.38 & 0.07 & 12.09 & -42.76 & 35.56 & 0.18 & 0.01 & 0.00 & 0.00 \\ 
  UBS & 0.07 & -0.04 & 4.15 & 0.18 & 15.79 & -47.25 & 35.19 & 0.12 & 0.01 & 0.00 & 0.00 \\ 
  UCG & 0.06 & -0.06 & 4.45 & 0.37 & 14.58 & -40.36 & 41.97 & 0.18 & 0.01 & 0.00 & 0.00 \\ 
  AXP & 0.04 & -0.03 & 3.83 & 0.47 & 13.39 & -31.41 & 30.27 & 0.00 & 0.01 & 0.00 & 0.00 \\ 
  BAC & 0.07 & -0.07 & 4.67 & 0.65 & 44.14 & -60.07 & 68.96 & 0.06 & 0.01 & 0.00 & 0.00 \\ 
  C & 0.06 & -0.09 & 4.33 & -0.16 & 29.03 & -52.11 & 40.12 & 0.12 & 0.01 & 0.00 & 0.00 \\ 
  GS & 0.04 & -0.09 & 4.31 & 0.39 & 43.43 & -57.55 & 64.15 & 0.06 & 0.01 & 0.00 & 0.00 \\ 
  JPM & 0.04 & -0.07 & 3.94 & 0.02 & 18.77 & -49.07 & 35.98 & 0.12 & 0.01 & 0.00 & 0.00 \\ 
  MS & 0.03 & -0.10 & 4.82 & -3.33 & 135.57 & -118.53 & 79.94 & 0.01 & 0.01 & 0.00 & 0.00 \\ 
  WFC & 0.07 & -0.05 & 4.03 & 1.15 & 25.13 & -41.46 & 54.58 & 0.15 & 0.01 & 0.00 & 0.00 \\ 
   \hline
\end{tabular}
\begin{tablenotes}
\item {\footnotesize\textbf{Notes:} The table reports the descriptive statistics for the log differences (in \%) of CDS spreads. 1stACF is the
             first-order autocorrelation,
             ADF($m$) is the $p$-value for the Augmented Dickey-Fuller test with $m$ lags,
the LB($k$) is  the $p$-value for the Ljung-Box test for  autocorrelation of $k$ lags,
and  ARCH($l$) is the $p$-value for the test for ARCH effects with $l$ lags. The data is  from January 1st, 2007  to May 1st, 2023.}
    \end{tablenotes}
      \end{threeparttable}  
}

\end{table}

\begin{figure}[!htbp]
\centering
\includegraphics[width=0.5\textwidth]{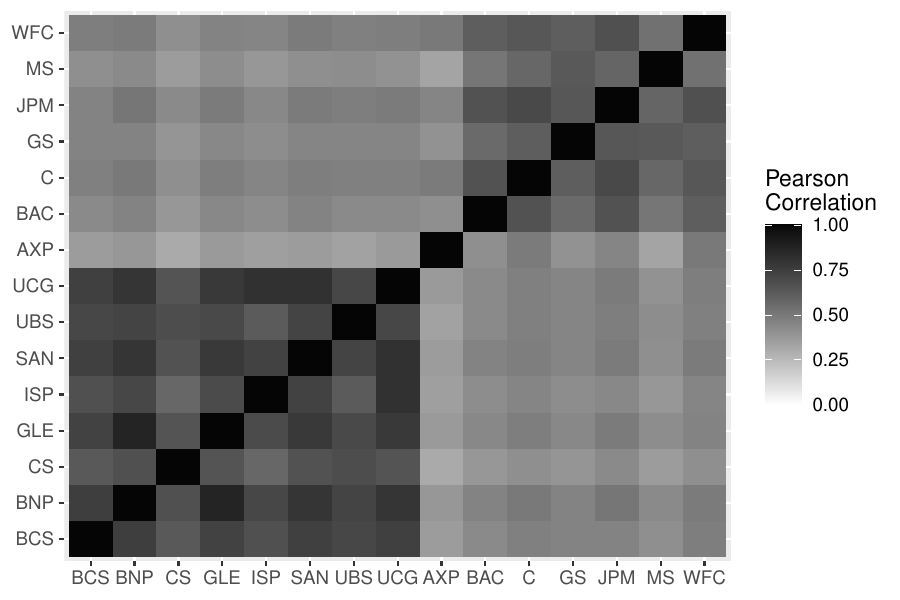}
\caption{Correlation matrix of CDS log-differences.}
\label{fig:cormat_lret}
\caption*{{\scriptsize The figure shows the sample Pearson correlation matrix between the log-differences of the CDS spreads. European banks are in the lower left quadrant, and U.S.\ banks are in the upper right quadrant.}}
\end{figure}

Finally, Figure \ref{fig:cormat_lret} plots the correlations between the log-differences. We can clearly see two groups: European banks are in the bottom left corner, meanwhile U.S.\ banks are in the right top corner. American Express Company (AXP) has the lowest correlation with any of the banks. The overall correlations are always positive and range from medium to very high, indicating a very high level of interconnectedness among the banks; more regional than global. European banks seem to show a stronger overall correlation than U.S. banks, indicating a more highly interconnected system.

\subsection{Marginals}
\label{sec:results_marginals}

Next, we model the log-differences of CDS spreads $s_{i,t}$ using an AR$\left(4\right)$-GJR-GARCH$\left(1,1\right)$ with the skew-$t$ errors, as described in Section \ref{sec:marginal}. All details related to the marginal modeling, such as the parameter estimation results, residual descriptive statistics, and plots are in the Online Appendix. Most of the banks have estimated positive AR$\left(1\right)$ and AR$\left(2\right)$ and negative AR$\left(3\right)$ and AR$\left(4\right)$ coefficients, similarly as in \citet{OhPatton2017}. The persistence parameter of the GJR-GARCH$\left(1,1\right)$ model indicates strong volatility persistence for all assets, and the asymmetric volatility response parameter $\gamma$ is negative in almost all cases, except for AXP, where it is statistically insignificant. In contrast to the ``conventional" financial assets, where positive log returns typically signal positive developments, positive values of CDS spread log-differences mean an increase in CDS spreads, which conveys adverse developments or negative news. Therefore, the estimated leverage parameter $\gamma$ has a negative sign: past negative log-differences are good news that decrease the current level of volatility. Finally, the estimated skewness and degrees of freedom parameters indicate that the error distribution is asymmetric and fat-tailed for all banks. The specified model is appropriate for the data, as the probability integral transforms of the log-differences result in uniformly distributed data, i.e.\ $F_{s_{i,t}}(s_{i,t})=u_{i,t}\overset{\text{iid}}{\sim} U(0,1)$, as shown in Online Appendix.

\subsection{In-sample results}
\label{sec:insample_results}

Next, we model the uniformly distributed data $\mathcal{U} = \{u_{i,t}:i = 1, \ldots d; t = 1,\ldots T\}$ using the one-factor, two-factor, nested-factor, bi-factor, and two-level truncated factor-vine copula models proposed in Section \ref{sec:Model_specification}. The \textit{ad hoc} grouping of the variables, necessary in the nested-factor and bi-factor copulas is based on the geographic location of the banks (Europe and U.S.). We have fitted the proposed models on the full sample as well as four approximately equal-length subsamples to account for a possible change of co-dependence structure over time. Table \ref{table:BIC} reports the in-sample Bayesian Information Criteria, while Online Appendix contains the results for the log likelihoods. Similarly, \citet{OhPatton2018} have also employed log likelihoods, AIC, and BIC criteria for the in-sample model evaluation.

\begin{table}[!htbp] 
\centering
\caption{In-sample Bayesian Information Criteria.} 
\label{table:BIC}
\scalebox{0.75}{
\begin{threeparttable}
\begin{tabular}{rccccc}
  \hline
 & Subsample 1 & Subsample 2 & Subsample 3 & Subsample 4 & Full sample \\ 
 & Jan 2007 - Dec 2010 & Jan 2011 - Dec 2014 & Jan 2015 - Dec 2019 & Jan 2020 - May 2023 & Jan 2007 - May 2023 \\ \hline
 \multicolumn{5}{l}{U.S.\ banks} \\
  One-factor &  -4200.99 & -5830.27 & -7006.16 & \sh -6170.19 & -22622.88 \\ 
  Two-factor & -4182.95 & -5803.62 & -6979.97 & -6098.57 & -22610.04 \\ 
  Factor-Vine & \sh -4455.18 & \sh -5834.02 & \sh -7017.22 & -6161.97 & \sh -22698.24\\ 
 \hline 
 \multicolumn{5}{l}{European banks} \\
  One-factor & -7359.31 & -8951.08 & -7829.39 & -6725.13 & -30878.72 \\ 
  Two-factor & -7407.53 & -8884.31 & -7873.32 & -6980.22 & -32215.13 \\ 
  Factor-Vine & \sh -7446.67 & \sh -9125.10 & \sh -8646.63 & \sh -7096.86 & \sh -32670.12 \\ 
\hline   
 \multicolumn{5}{l}{U.S.\ and European banks} \\
  One-factor & -9175.17 & -11433.56 & -10301.33 & -10392.73 & -41131.21 \\ 
  Two-factor & -11744.15 & -14924.44 & -14964.80 & -13293.54 & -54773.96 \\ 
  Bi-factor & \sh -12016.55 & \sh -15322.08 & \sh -15809.10 & \sh -13429.70 & \sh -56212.63 \\ 
  Nested-factor & -11761.48 & -15202.06 & -15191.11 & -13472.43 & -54336.74 \\ 
  Factor-Vine & -11563.51 & -14732.43 & -15498.63 & -13346.46 & -53731.25 \\  \hline  
\end{tabular}
\begin{tablenotes}
\item {\footnotesize\textbf{Notes:} The table reports the in-sample Bayesian Information Criteria (BIC) for four subperiods and the full sample.}
    \end{tablenotes}
      \end{threeparttable}  
}

\end{table}

The first four columns of Table \ref{table:BIC} contain the BICs for the four subperiods, meanwhile, the last column reports the overall BIC for the full sample. For both regions individually, the factor-vine copula is decidedly the best, indicating that both contagion channels are at work: systematic and idiosyncratic.  Nonetheless, there is one exception. For the U.S.\ in the subsample 4 (January 2020 - May 2023), a period that includes COVID-19, 2022 interest rate spikes, early 2023 banking panic, the best is the one-factor copula. From the economic perspective, this means that during this specific period, the interconectedness of the U.S.\ banking sector can be explained due to some common latent factor rather than the interactions between banks. Finally, in considering both regions together, the structure that describes the co-dependence between U.S.\ and European banks the best is the bi-factor model. Such specification permits for a common latent global factor, and two regional latent factors to govern the co-dependence in the banking sector. The above results provide some interesting insights: on a regional level, banking sector contagion has two channels, systematic and idiosyncratic (inter-institutional). However, on the global level, the banking sector is mostly affected by global and regional common factors, and not by individual banks, i.e.\ contagion is mostly due to systematic factors. The results are consistent with the findings in the related literature. For example, \citet{Ballester2016} has studied financial contagion between European and U.S.\ banks using the CDS spread data. By employing a generalized VAR approach, they show that the systematic contagion component is always greater than the idiosyncratic component, emphasizing the importance of common factors in the risk spillover transmission. 

For all models, the vast majority of factor-unit links are $t$ copulas, indicating fat-tailed but symmetric dependence between the latent factor(s) and individual banks. In other words, the systematic contagion channel is overall homogeneous and best described using $t$ distributions. The degrees of freedom parameter can vary from 2.2 to 41.1 for different models, different banks, and different subsamples. For the factor-vine copula, even though the factor-unit links are mostly $t$, the unit-unit links at the second tree level are more heterogeneous, and depend on the units (banks) as well as the time period. As an example, Figure~\ref{fig:fvcopP4emp} draws the second tree level for the factor-vine copula for the last subperiod and the full sample. For the last subperiod (top row), in the U.S., the selected C-vine structure assigns three links to Wells-Fargo bank with $t$, Frank, and Gumbel copulas. In Europe, the Italian UniCredit (UCG) bank has as many as five links, with many different types of links. Bottom row presents a similar plot, but for the entire sample. For the U.S., the Bank of America (BAC) now has the most links, all Frank copulas (symmetric and fat-tailed). In Europe, the UniCredit bank remains with the highest number of links. Interestingly, the link structure is more stable for Europe than for the U.S.\ across both subsamples. In other words, for the U.S, most of the co-dependence is captured in the first level via the latent factor, and the remaining co-dependence is weak and more difficult to identify, resulting in different vine copula structures for each subperiod. On the other hand, in the European region, the latent factor at the first tree level captures only a part of the co-dependence, and the remaining links at the second level are relatively strong and retain similar structures across different subsamples. 

\begin{figure}[!htbp] 
    \begin{center}
    \begin{minipage}{.45\textwidth}
\begin{tikzpicture}[thick, scale=0.7, every node/.style={scale=0.7}]
    \node (sensor) [sensor] (b1) at (0,0) {\scriptsize $AXP$};
     \node (sensor) [sensor] (b2) at (6,0) {\scriptsize $BAC$};
     \node (sensor) [sensor] (b3) at (2,0) {\scriptsize $C$};
     \node (sensor) [sensor] (b4) at (4,1.5) {\scriptsize $GS$};
     \node (sensor) [sensor] (b5) at (8,1.5) {\scriptsize $JPM$};
     \node (sensor) [sensor] (b6) at (8,-1.5) {\scriptsize $MS$};
     \node (sensor) [sensor] (b7) at (4,0) {\scriptsize $WFC$};

     \draw[-, very thick] (b1) to  node[midway,above] {G} (b3) ;
     \draw[-, very thick] (b2) to node[midway,above] {$t$} (b6) ;
     \draw[-, very thick] (b2) to node[midway,above] {F} (b5) ;
     \draw[-, very thick] (b2) to node[midway,above] {G} (b7) ;
     \draw[-, very thick] (b3) to node[midway,above] {$t$} (b7) ;
     \draw[-, very thick] (b4) to node[midway,left] {F} (b7) ;
\end{tikzpicture}

\end{minipage}
\begin{minipage}{.45\textwidth}

\begin{tikzpicture}[thick, scale=0.7, every node/.style={scale=0.7}]
    \node (sensor) [sensor] (b1) at (0,0) {\scriptsize $BCS$};
     \node (sensor) [sensor] (b2) at (2,1.5) {\scriptsize $BNP$};
     \node (sensor) [sensor] (b3) at (2,-1.5) {\scriptsize $CS$};
     \node (sensor) [sensor] (b4) at (6,1.5) {\scriptsize $GLE$};
     \node (sensor) [sensor] (b5) at (0,1.5) {\scriptsize $ISP$};
     \node (sensor) [sensor] (b6) at (6,-1.5) {\scriptsize $SAN$};
     \node (sensor) [sensor] (b7) at (0,-1.5) {\scriptsize $UBS$};
     \node (sensor) [sensor] (b8) at (4,0) {\scriptsize $UCG$};

     \draw[-, very thick] (b2) to  node[midway,above] {F} (b5) ;
     \draw[-, very thick] (b1) to node[midway,above] {C} (b8) ;
     \draw[-, very thick] (b3) to node[midway,above] {$t$} (b7) ;
     \draw[-, very thick] (b2) to node[midway,above] {F} (b8) ;
     \draw[-, very thick] (b3) to node[midway,above] {$t$} (b8) ;
     \draw[-, very thick] (b4) to node[midway,left] {F} (b8) ;
     \draw[-, very thick] (b6) to node[midway,left] {C} (b8) ;
\end{tikzpicture}

\end{minipage}
\end{center}

\begin{center}
    \begin{minipage}{.45\textwidth}
\begin{tikzpicture}[thick, scale=0.6, every node/.style={scale=0.6}]
    \node (sensor) [sensor] (b1) at (10,1.5) {\scriptsize $AXP$};
     \node (sensor) [sensor] (b2) at (8,0) {\scriptsize $BAC$};
     \node (sensor) [sensor] (b3) at (4,0) {\scriptsize $C$};
     \node (sensor) [sensor] (b4) at (6,0) {\scriptsize $GS$};
     \node (sensor) [sensor] (b5) at (2,0) {\scriptsize $JPM$};
     \node (sensor) [sensor] (b6) at (10,-1.5) {\scriptsize $MS$};
     \node (sensor) [sensor] (b7) at (0,0) {\scriptsize $WFC$};

     \draw[-, very thick] (b1) to  node[midway,above] {F} (b2) ;
     \draw[-, very thick] (b2) to node[midway,above] {F} (b6) ;
     \draw[-, very thick] (b2) to node[midway,above] {F} (b4) ;
     \draw[-, very thick] (b3) to node[midway,above] {F} (b4) ;
     \draw[-, very thick] (b3) to node[midway,above] {$t$} (b5) ;
     \draw[-, very thick] (b5) to node[midway,above] {$t$} (b7) ;
\end{tikzpicture}

\end{minipage}
\begin{minipage}{.45\textwidth}

\begin{tikzpicture}[thick, scale=0.7, every node/.style={scale=0.7}]
    \node (sensor) [sensor] (b1) at (0,0) {\scriptsize $BCS$};
     \node (sensor) [sensor] (b2) at (8,1.5) {\scriptsize $BNP$};
     \node (sensor) [sensor] (b3) at (2,-1.5) {\scriptsize $CS$};
     \node (sensor) [sensor] (b4) at (6,1.5) {\scriptsize $GLE$};
     \node (sensor) [sensor] (b5) at (4,1.5) {\scriptsize $ISP$};
     \node (sensor) [sensor] (b6) at (6,-1.5) {\scriptsize $SAN$};
     \node (sensor) [sensor] (b7) at (0,-1.5) {\scriptsize $UBS$};
     \node (sensor) [sensor] (b8) at (4,0) {\scriptsize $UCG$};

     \draw[-, very thick] (b2) to  node[midway,above] {F} (b4) ;
     \draw[-, very thick] (b1) to node[midway,above] {C} (b8) ;
     \draw[-, very thick] (b3) to node[midway,above] {$t$} (b7) ;
     \draw[-, very thick] (b4) to node[midway,above] {F} (b5) ;
     \draw[-, very thick] (b3) to node[midway,above] {$t$} (b8) ;
     \draw[-, very thick] (b4) to node[midway,left] {F} (b8) ;
     \draw[-, very thick] (b6) to node[midway,left] {C} (b8) ;
\end{tikzpicture}

\end{minipage}
\end{center}
    \caption{Second level of the truncated factor-vine copula.}
    \label{fig:fvcopP4emp}
    \caption*{{\scriptsize The figure shows the detailed second tree level of factor-vine copula for the U.S.\ (on the left) and European (on the right) banks. Top row is for the last subsample: from January 2020 to May 2023, bottom row is for the entire period: from January 2007 to May 2023.}}
\end{figure}

\subsection{Out-of-sample results} \label{sec:oos_results}

For the out-of-sample forecasting exercise, we retain the last 1000 observations, from February 28th 2019 till May 1st, 2023. We then roll the dataset in 20-day increments (approximately one month), re-estimate all the models, and predict the co-dependence structure for the next 20-day period. In order to evaluate the out-of-sample forecasting performance of individual models we calculate the negative log predictive score (LPS), the negative conditional likelihood score (CdL), and the variogram score (VarS). The LPS takes into account the goodness of fit of the entire predictive distribution and is one of the most commonly used measures to evaluate the predictive model performance. The CdL, on the other hand, allows to evaluate a certain region of the predictive distribution, for example, the upper tail. The upper tail is defined as a $d$-dimensional quadrant where the CDS log-differences for all the banks fall within the top 50th percentile. This typically covers approximately 10\% of the data, depending on the time period. Note that contrary to the ``conventional" assets, the extreme negative events of the CDS spread log-differences happen in the upper tail, and not in the lower tail. Finally, the VarS is based on pairwise differences of the components of the multivariate forecast. This way we can evaluate point-wise forecast accuracy. For details concerning the calculation of each of the scores and their properties, refer to \citet{Bjerregard2021}. The results for LPS, CdL, and VarS for the proposed models are reported in Table \ref{table:oos}.

\begin{table}[!htbp] 
\centering
\caption{Out-of-sample forecast results.} 
\label{table:oos}
\scalebox{0.8}{
\begin{threeparttable}
\begin{tabular}{rccccc}
  \hline
& One-factor & Two-factor & Bi-factor & Nested-factor & Factor-Vine \\ \hline
\multicolumn{6}{l}{(a) Negative log predictive scores } \\
  US banks & -2874.50 & -2884.82 & - & - & \sh -2892.41 \\ 
  Euro banks & -3165.57 & -3254.41 & - & - & \sh -3307.69 \\ 
  US and Euro banks & -4932.64 & -6196.65 & \sh -6365.17 & -6313.78 & -6127.93 \\  \hline  
  \multicolumn{6}{l}{(b) Negative conditional likelihood predicted scores } \\
US banks & -1532.82 & \sh -1537.80 & - & - & -1533.43 \\ 
  Euro banks & -1755.38 & \sh -1799.65 & - & - & -1788.37 \\ 
  US and Euro banks & -1910.98 & -2036.85 & -2032.17 & -1977.84 & \sh -2037.89 \\ 
   \hline
    \multicolumn{6}{l}{(c) Variogram score ($p=0.5$) } \\
US banks & 811.76 & 811.84 & - & - & \sh 809.90 \\ 
  Euro banks & 1020.67 & 1017.73 & - & - & \sh 1016.08 \\ 
  US and Euro banks & 4324.11 & 4323.60 & \sh 4264.03 & 4430.45 & 4272.60 \\   \hline
\end{tabular}
\begin{tablenotes}
\item {\footnotesize\textbf{Notes:} The table reports the sum of the negative log predictive scores (LPS), negative conditional likelihood scores (CdL),  and variogram scores (VarS) for the last 1000 observations (from February 28th, 2019 to May 1st, 2023, both inclusive). The scores are calculated by rolling the sample every 20 days, re-estimating the models, and predicting the co-dependence structure for the next 20 days. Lower numbers mean better. The parameter $p=0.5$ next to the VarS indicates the order of the score, see the discussion in \cite{Bjerregard2021}.}
    \end{tablenotes}
      \end{threeparttable}  
}

\end{table}

In terms of LPS and VarS, the out-of-sample forecasting results are consistent across the two scoring rules. Similar to the in-sample results, the banking sector co-dependence is best described via factor-vine copula if we consider both regions individually. In other words, the financial contagion has systematic and idiosyncratic sources in regional financial systems. On the other hand, if we consider both banking systems jointly, then the bi-factor copula provides the best out-of-sample forecasting results. This means that both financial systems are governed by some common global factor and two individual regional factors. In this case, the bank-to-bank (idiosyncratic) contagion is not as important. Interestingly, the CdL metric provides alternative model ordering: if we consider the upper tail of the distribution only, then for both regions the two-factor copula is the best, meanwhile for both regions jointly the factor-vine copula provides the smallest value of the CdL. In other words, to account for the behavior in the upper tail, a multi-source systematic contagion channel is prevalent within each region individually. Furthermore, when considering the upper tail of the European and U.S. banking systems collectively, both global and local contagion channels are in operation.

\subsection{Measuring systematic risk}\label{sec:systrisk_results}

Modeling and forecasting CDS series allows to model and forecast instances of financial distress, which are defined as events where the implied probability of default exceeds a certain threshold. Instead of relying solely on the implied probability of default, we focus on the directly observable CDS series to identify periods of distress. A similar approach was employed by \cite{OhPatton2018}, for example.
The CDS prediction is carried out by simulating 10,000 samples of the $d$-variate variable $u_t$ and applying the inverse probability integral transform (the quantile function) resulting from the forecasted AR($p$)-GJR-GARCH(1,1)-skew-$t$ model. This procedure is repeated for each forecast instance $t$ and for each asset $i$.

This enables us to calculate four measures of systematic risk, namely the individual Probability of Distress ($PD$) \citep{OhPatton2018}, the Joint Probability of Distress ($JPD$) \citep{Segoviano2009, Lucas2014, OhPatton2018, Krupskii2020}, the Expected Proportion in Distress ($EPD$) \citep{Hartmann2005,OhPatton2018} and the Expected Shortfall ($ES$) \citep{OhPatton2017}.

Distress happens when the observed CDS spread is larger than some threshold. The distress indicator for bank $i$, for $i=1,\hdots,d$ can be defined as $D_{i,t+20}=\mathbf{1}\{S_{i,t+20}>C_{i,t+20}\}$. Hence, the distress indicator $D$ will be equal to one if the 20-step-ahead predicted CDS spread $S_{i,t+20}$ is larger than some threshold $C$. The threshold can be calculated as the upper $95$ (or $99$) percentile of the historical CDS spreads. Since in the forecasting exercise we have a rolling sample, the threshold $C$ is time-varying. Therefore, $C$ is recalculated as the upper $95$ percentile of the last 1000 observations of the CDS spreads. Then the systematic risk measures are calculated as follows:
\begin{align*}
    PD_{i,t}&={\Pr}_{t}\left[D_{i,t+20}=1\right],\\
    JPD_{k,t}&={\Pr}_{t}\left[\left(\frac{1}{d}\sum_{i=1}^d D_{i,t+20}\right)\geq \frac{k}{d}\right],\\
    EPD_{i,t} &= \textrm{E}_t\left[\frac{1}{d} \sum_{j=1}^d D_{j,t+20}\:  \vline \: D_{i,t+20}=1 \right],\\
    ES_{i,t} & = \textrm{E}_{t}\left[S_{i,t+20}\left|\left( \sum_{j=1}^d \mathbf{1}\{S_{j,t+20}>C_{j,t+20}\}\right)>1\right.\right].
\end{align*}
First, $PD$ measures the probability of each individual bank being in distress in the next 20 days. Since at each instance $t$ for each asset $i$ we simulate 10,000 scenarios for the spreads $S_{i,t+20}$, this probability is estimated as a sample proportion of the spreads that reach beyond the threshold. Second, the $JPD$ measures the probability that a certain proportion of banks will be in distress. This probability should be decreasing as $k$ increases since it is less and less likely for more and more banks to be in distress at the same time. Third, the $EPD$ presents an expected proportion of banks in distress when bank $i$ has suffered distress. It might provide some insights into individual banks' relative importance in the system and its effect on the rest of the banks. Finally, the $ES$ metric calculates the expected CDS spread of bank $i$ given that $2$ or more banks have CDS spreads above some threshold $C$. 

\begin{figure}[htbp]
  \centering
  \begin{subfigure}[b]{0.5\linewidth}
    \includegraphics[width=\linewidth]{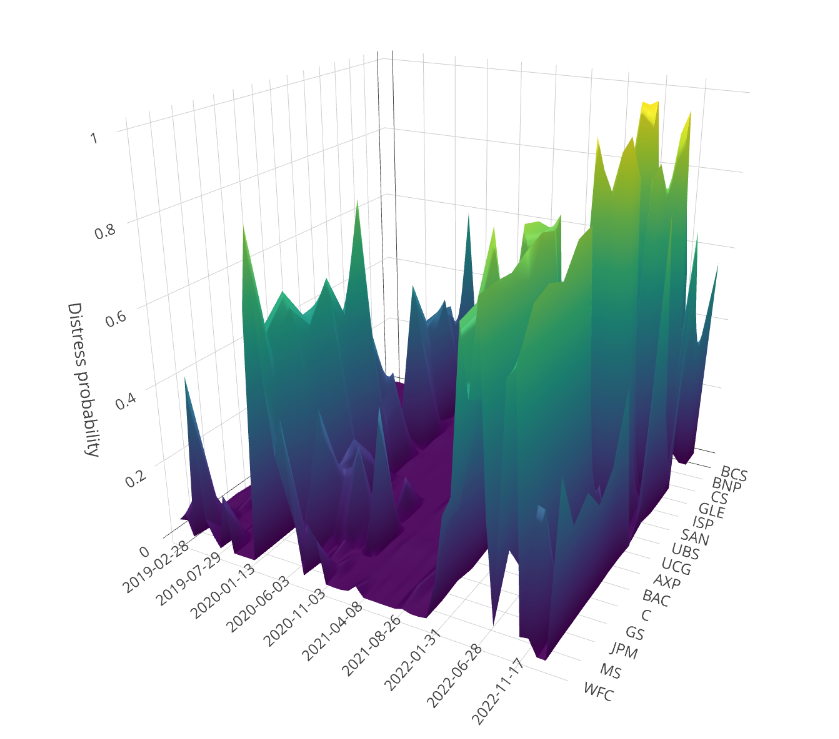}
    \caption{Individual probability of distress.}
    \label{fig:plot1}
  \end{subfigure}
  \hspace{-0.06\linewidth} 
  \begin{subfigure}[b]{0.5\linewidth}
    \includegraphics[width=\linewidth]{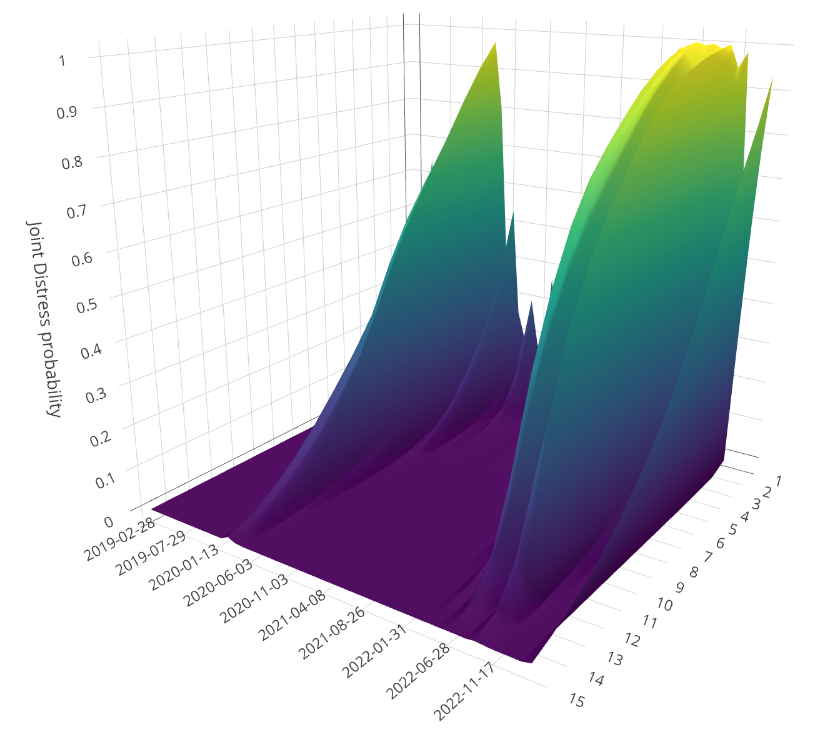}
    \caption{Joint probabilities of distress.}
    \label{fig:plot2}
  \end{subfigure}
  
  \begin{subfigure}[b]{0.5\linewidth}
    \includegraphics[width=\linewidth]{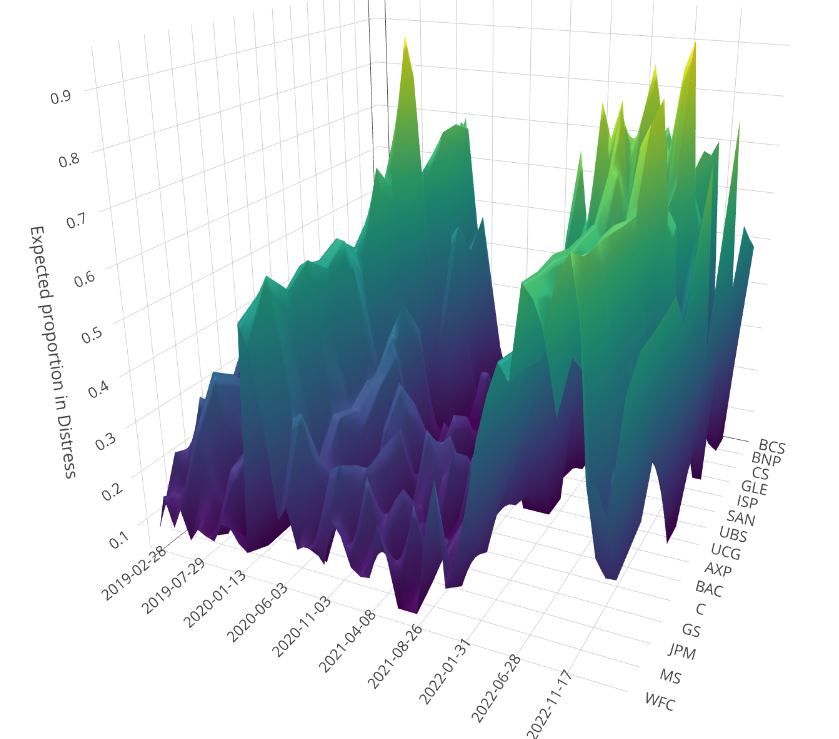}
    \caption{Expected proportion in distress.}
    \label{fig:plot3}
  \end{subfigure}
  \hspace{-0.06\linewidth} 
  \begin{subfigure}[b]{0.5\linewidth}
    \includegraphics[width=\linewidth]{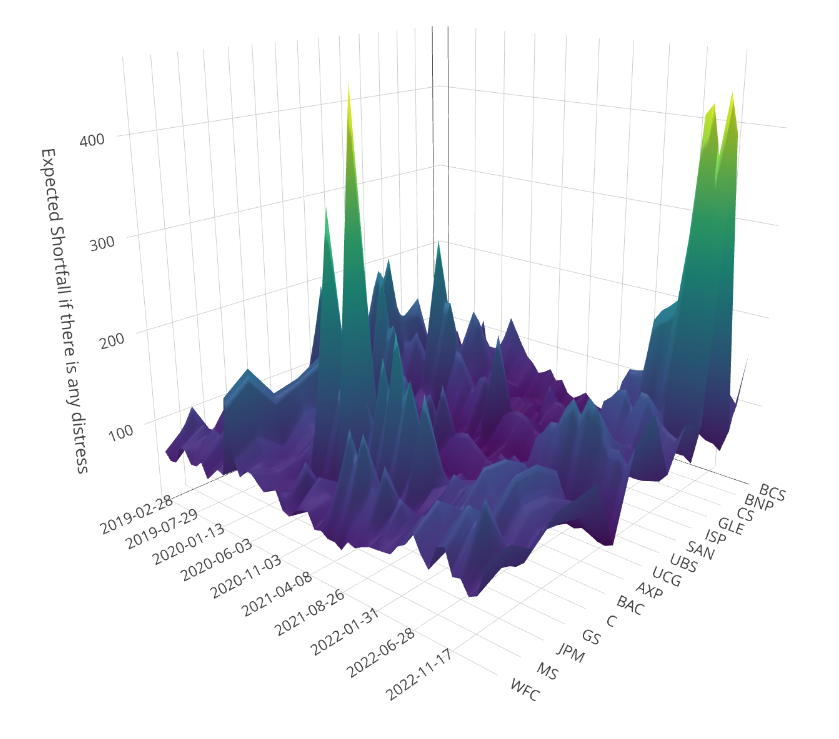}
    \caption{Expected shortfall.}
    \label{fig:plot4}
  \end{subfigure}
  
  \caption{Forecasts of systematic risk.}
  \label{fig:composite}
  \caption*{{\scriptsize The figure shows the forecast measures of systematic risk - $PD$, $JPD$, $EPD$, and $ES$ - in the next 20 days for the out-of-sample period from February 28th, 2019
            to March 29th, 2023.}}
\end{figure}

Figure \ref{fig:composite} summarizes the results for all banks for the out-of-sample period. Please note that the 3D plots offer a general overview of the trends in systematic risk measures. For precise values of a specific risk measure for a particular bank on a given day, refer to the detailed tables in the Online Appendix. Also, the presented results are for the entire system as a whole, i.e.\ using the bi-factor copula model. All detailed tables for the region-wise (Europe and the U.S.) systematic risk measures are available in the Online Appendix as well.

First, from sub-figure \ref{fig:plot1} we can observe an overall increase in the individual distress probabilities (PD) in March - April 2020. The increase was universal for all banks in the system, to as high as 0.83 for WFC. This dramatic increase in the distress probability can be attributed to the COVID-19 pandemic. \citet{Stepankova2023} have documented the increase in the probability of default in the banking sector during the COVID-19 pandemic, and \citet{Duan2021} have found that the pandemic has increased systemic risk in the banking sector across as many as 64 countries. Another universal peak can be seen in the year 2022 and the beginning of 2023, with a sharp drop in August 2022. This persistent high probability of distress throughout the year can be attributed to the hike in the interest rates: only in 2022 the Fed raised interest rates five times and ECB - three times.  As noted in a report by the \citet{FSOC2013}, a fast spike in yields can affect financial stability. For example, if a bank has a large portion of its assets locked in long-term bonds, the market value of those bonds goes down when interest rates start going up. If at the same time, a large number of depositors withdraw their funds due to concerns about the bank's solvency or liquidity (a phenomenon called a bank run), this might lead to a liquidity crunch and, in an extreme case, a bank’s failure \citep{Lyocsa2023}. This indeed played an important part in the collapse of three regional U.S.\ banks (the Silicon Valley, Signature, and First Republic banks) in early 2023 \citep{Lyocsa2023}, which also triggered the downfall of the already vulnerable Credit Suisse. These events are visible in the plot as the peak of the probability of distress at the end of the sample. 


Next, sub-figure \ref{fig:plot2} draws the Joint Probability of Distress (JPD) series that represents the probability that at a given moment in time, at least $d$ banks will be in distress, for $d=1,\ldots,15$. Similarly to the previous plot, there are two major turbulent periods, namely the COVID-19 pandemic and the interest rate hikes in 2022. During those periods, the probability that at least one bank will be in distress reaches almost one. Same as in the individual distress probabilities series, there is a sharp drop in the $JPD$ in August 2022, which has no apparent explanation. Finally, the 2022 - March 2023 peak is also visible at the end of the period.  


Sub-figure \ref{fig:plot3} presents the Expected Proportion in Distress (EPD), which permits to evaluate the individual banks' relative importance in the system. For example, on February 28th, 2019 the $EPD$ value for Barclay's bank (BCS) is 0.13, which means that if in 20-days-ahead Barclay's bank suffers distress, then 13\% of all banks in the system will also suffer distress. Overall, it seems that U.S.\ banks' relative importance in the overall system is of smaller magnitude as compared to the European banks, at least for the selected sample. Interestingly, during the beginning of the COVID-19 pandemic, the Intesa Sanpaolo bank's $EPD$ skyrocketed from 0.07 to as high as 0.89. Italy was the European country where COVID-19 struck the strongest: it had the highest death toll and it was the first one to go into lockdown. Therefore, if Italy's largest bank, Intesa Sanpaolo, were to suffer distress during that period, that would have sent a very strong signal to the rest of the countries about the possible adverse effects of the COVID-19 pandemic on the banking sector. The figure also gives a good approximation of the periods for ``heightened sensitivity". For example, we can observe a universal increase in the $EPD$ values during periods of financial distress: the COVID-19 pandemic, the 2022 interest rate hikes, and the 2023 collapse of some European and U.S.\ banks.


Finally, sub-figure \ref{fig:plot4} summarizes the results for the Expected Shortfall (ES) of CDS spreads when 2 or more banks are in distress. For example, during the COVID-19 pandemic (the very beginning of 2021), Goldman Sachs bank had the predicted $ES$ value of 459 bps, given that two or more other banks in the system were to suffer distress. This indicates a high vulnerability of the Goldman Sachs bank during this specific period. In particular, for that specific time, Goldman Sachs's probability of distress is estimated to be 0.37, which can be attributed to a sudden 30\% daily increase in the CDS spreads (from 52 to 71 bps). Finally, at the end of the sample, Credit Suisse exhibits expected CDS spreads as high as 400, which closely resembles the actual observed CDS data for this specific bank.


\section{Conclusion} 
\label{sec:conclusion}

In this paper, we employ five different models using multi-factor, structured factor, and factor-vine copulas to forecast both the joint and conditional probabilities of distress for a sample of banks. We employ the Variational Bayes estimation method to approximate the posterior distributions of both latent factors and copula model parameters. We also incorporate an automatic procedure to select the best bivariate links among the nodes. We find that in-sample, for U.S.\ and Europe individually, the best fitting model is the factor-vine copula. In other words, there are two financial contagion channels at work: systematic (governed by some latent factor at the first level) and idiosyncratic (governed by vine copulas at the second level). Most of the factor-unit links at the first level are $t$ copulas indicating fat-tailed and symmetric dependence structure, meanwhile, the unit-unit links at the second level are of all types: asymmetric and/or fat-tailed, depending on the banks and the time period. Moreover, if we consider U.S.\ and European banking systems jointly, the interconnectedness between the CDS spread log-differences is best captured using the bi-factor copula. This shows the existence of one common global factor and two regional factors, pointing to the dominance of the systematic contagion channel. Using the best-fitting copula models we then forecast a variety of risk measures that capture the individual, joint, and conditional probabilities of distress. The findings closely coincide with the observed historical data: there are clear peaks in probabilities of distress during turbulent periods, such as the COVID-19 pandemic, 2022 interest rate hikes, and the 2023 banking panic. Also, the model captures the increase in the probability of distress for certain banks in the system, such as Credit Suisse for example. 


\section*{Acknowledgments}
The second author is partially supported by the grants PID2022-138289NB-I00 (\textit{Modeling and forecasting financial returns and macroeconomic variables in data-rich environments: new methods and applications}) and PID2020-113192GB-I00/AEI/10.13039/501100011033 (\textit{Mathematical Visualization: Foundations, Algorithms and Applications}) from the Spanish State Research Agency (Ministerio de Ciencia e Innovaci\'on). The third and fourth authors are supported by the grant PID2022-138114NB-I00/AEI/10.13039/501100011033 (\textit{High dimensional dependence modeling}) from the Spanish State Research Agency (Ministerio de Ciencia e Innovaci\'on).

\noindent\textbf{Declaration of interest}: None 

\appendix
\begin{flushleft}
{\Large \textbf{Appendix}}
\end{flushleft}

\section{Factor copulas and factor-vine copulas}\label{app:fact}
\subsection{One-factor copula models}

In the one-factor copula model, the variables $U_1, \ldots, U_d $ are conditionally independent given the latent variable, $V_0$, as shown in \cite{Krupskii2013}. We have that the conditional distribution function of $U_i$ given $V_0$ is given by,
 $$ F_{U_i|V_0}(u_i | v_0; \boldsymbol{\theta_{0i}}) = \frac{\partial  F_{U_i,V_0}(u_i , v_0; \boldsymbol{\theta_{0i}}) }{ \partial v_0  } = \frac{\partial  C_{U_i,V_0}(u_i , v_0; \boldsymbol{\theta_{0i}}) }{ \partial v_0  },$$ 
where $F_{U_i,V_0}$ is the joint CDF of $U_i$ and $V_0$, and $C_{U_i, V_0} (u_i, v_0; \boldsymbol{\theta_{0i}})$ is the bivariate copula distribution function parameterized by $\boldsymbol{\theta_{0i}}$ as the vector of parameters of the bivariate copula. 
Then, the conditional copula density and the integrated copula density can be derived respectively as,
\begin{equation}\label{eq:onefcop}
\begin{aligned}
p(u_1,\ldots,u_d  | v_0; \boldsymbol{\theta}) &= \frac{\partial F(u_1,\ldots,u_d | v_0; \boldsymbol{\theta}) }{ \partial u_1 \ldots \partial u_d  } = \prod_{i = 1}^d \frac{\partial  F_{U_i|V_0}(u_i | v_0; \boldsymbol{\theta}) }{ \partial u_i  } = \prod_{i = 1}^d \frac{\partial  C_{U_i,V_0}(u_i , v_0; \boldsymbol{\theta}) }{ \partial u_i \partial v_0  }  \\
& = \prod_{i = 1}^d c_{U_i , V_0}(u_i , v_0 ; \boldsymbol{\theta_{0i}}), 
\end{aligned}
\end{equation}
and,
\begin{equation}
p(u_1,\ldots,u_d ; \boldsymbol{\theta} ) = \int_{0}^1 p(u_1,\ldots,u_d  | v_0; \boldsymbol{\theta}) dv_0 = \int_{0}^1 \prod_{i = 1}^d c_{U_i , V_0}(u_i , v_0 ; \boldsymbol{\theta_{0i}}) dv_0,
\end{equation}
where $\boldsymbol{\theta} = [\boldsymbol{\theta_{01}^{'}}, \ldots, \boldsymbol{\theta_{0d}^{'}}]^{'}$ is the vector of copula parameters.

\subsection{Two-factor copula models}

Similarly, \cite{Krupskii2013} compute the conditional copula density and the integrated copula density of the two-factor copula model respectively as,
\begin{equation}
\begin{aligned}
p\left(u_1,\ldots,u_d\mid v_0, v_1 ; \boldsymbol{\theta}\right) & =\prod_{i=1}^{d} p\left(u_{i} \mid v_0, v_1 ; \boldsymbol{\theta}\right) \\ & = \prod_{i=1}^{d} c_{U_{i}, V_1 \mid V_0}\left(u_{i \mid v_0}, v_1 ;\boldsymbol{\theta_{1i}} \right) c_{U_{i}, V_0}\left(u_{i}, v_0 ; \boldsymbol{\theta_{0i}}\right), \end{aligned}
\end{equation}
and
\begin{equation}
p\left(u_1,\ldots,u_d;\boldsymbol{\theta}\right)  =\int_{0}^{1}\int_{0}^{1} p\left(u_1,\ldots,u_d \mid v_0,v_1;\boldsymbol{\theta}\right) p(v_1) p(v_0) dv_1 dv_0,
\end{equation}
where $u_{i|v_{0}}=F_{U_i|V_0}(u_i|v_0)$, and $\boldsymbol{\theta} = [\boldsymbol{\theta_{01}^{'}}, \ldots, \boldsymbol{\theta_{0d}^{'}}, \boldsymbol{\theta_{11}^{'}}, \ldots, \boldsymbol{\theta_{1d}^{'}}]^{'}$ is the set of copula parameters. The joint copula distribution and density require a high dimensional integral which makes it difficult to consider more than three latent variables, see \citet{Joe2014}. 

\subsection{Structured factor copula models}
The nested factor copula model defines a hierarchical dependence structure from a common root variable. 
The conditional density function for the nested factor copulas is derived following \cite{Krupskii2015},
\begin{equation}
\begin{aligned}
p(u_1,\ldots,u_d  , v_1, \ldots, v_G | v_0 ; \boldsymbol{\theta} ) &= p(u_1,\ldots,u_d  | v_1, \ldots, v_G; \boldsymbol{\theta}) p(v_1, \ldots, v_G | v_0; \boldsymbol{\theta}),
\end{aligned}
\end{equation}
where, similar to the one-factor copula, the conditional copula density for the observables is, 
\begin{equation*}
\begin{aligned}
p(u_1,\ldots,u_d  | v_1, \ldots, v_G; \boldsymbol{\theta}) = \prod_{g = 1}^G \prod_{i_g = 1}^{d_g} c_{U_{i_g} , V_g}(u_{i_g} , v_g ; \boldsymbol{\theta_{gi_{g}}}),\\
\end{aligned}
\end{equation*}
and for the latent factor is,
\begin{equation*}
\begin{aligned}
p(v_1, \ldots, v_G | v_0; \boldsymbol{\theta}) = \prod_{g = 1}^G c_{V_{g} , V_0}(v_{g} , v_0 ; \boldsymbol{\theta_{0g}}). \\
\end{aligned}
\end{equation*}
where $c_{U_{i_g} , V_g}(u_{i_g} , v_g ; \boldsymbol{\theta_{gi_{g}}})$ is the bivariate copula density of $U_{i_g}$ and $V_g$ for each group member $i_g$ in group $g$ with $d_g$ members, for $g = 1, \ldots, G$; and $c_{V_{g} , V_0}(v_{g} , v_0 ; \boldsymbol{\theta_{0g}})$  is the bivariate copula density of $V_{g}$ and $V_0$; and 
$\boldsymbol{\theta} = [\boldsymbol{\theta_{01}^{'}}, \ldots, \boldsymbol{\theta_{0G}^{'}}, \boldsymbol{\theta_{11}^{'}}, \ldots, \boldsymbol{\theta_{G d_G}^{'}} ]^{'}$ is the set of copula parameters.

Similar to the two-factor copula, the conditional density function for the bi-factor copula model is given by,
\begin{equation}
\begin{aligned}
p(u_1,\ldots,u_d | v_0, \ldots, v_G; \boldsymbol{\theta}) &= \prod_{g=1}^{G} \prod_{i_g=1}^{d_g} p(u_{i_g} | v_0, v_g; \boldsymbol{\theta}) \\
& = \prod_{g=1}^{G} \prod_{i_g=1}^{d_g} c_{U_{i_g}, V_g|V_0}(u_{i_{g}|v_0} , v_g ; \boldsymbol{\theta_{gi_{g}}}) c_{U_{i_g} , V_0}(u_{i_g} , v_0 ; \boldsymbol{\theta_{0i_g}}), \\
\end{aligned}
\end{equation}
where $u_{i_g|v_{0}}=F_{U_{i_g}|V_0}(u_{i_g}|v_0)$ and $\boldsymbol{\theta} = [\boldsymbol{\theta_{01}^{'}}, \ldots, \boldsymbol{\theta_{0d}^{'}}, \boldsymbol{\theta_{11}^{'}}, \ldots, \boldsymbol{\theta_{G d_G}^{'}} ]^{'}$ is the vector of copula parameters.

\subsection{Factor-vine copula models}
 We follow \cite{Kurowicka06}, \cite{Kurowicka06}, \cite{Brechmann2012}, \cite{Krupskii2013} to derive the joint density of the truncated factor vine copula model,
\begin{equation} \label{eq:denefcop}
\begin{aligned}
p(v_0, u_1, \ldots, u_d ; \boldsymbol{\theta}) = \prod_{i = 0}^{K} \prod_{e(j,k) \in E_i} c_{j, k | D(e)} ( u_{j | D(e)} ,u_{k | D(e)}   ; \boldsymbol{\theta})
\end{aligned}
\end{equation}
where $u_{j | D(e)} = F_{U_j | U_{D(e)} }(u_{j} | u_{D(e)})$, $u_{k | D(e)} = F_{U_k | U_{D(e)} }(u_{k} | u_{D(e)})$ and $u_{D(e)}$ is a sub-vector of $u$ with the conditioning set $D(e)$. The pseudo $u_{j | D(e)}$ can be computed recursively using the formula derived in \cite{Aas2009}. 
In order to calculate the copula density of observable variables, one can take the integral of the conditional copula density over the latent space,
\begin{equation} \label{eq:copden}
\begin{aligned}
p(u_1, \ldots, u_d ; \boldsymbol{\theta}) = \int_{0}^1 \prod_{i = 0}^{K} \prod_{e \in E_i} c_{j, k | D(e)} ( u_{j | D(e)} ,u_{k | D(e)} ; \boldsymbol{\theta}) dv_0.
\end{aligned}
\end{equation}
For example, the conditional density of a truncated factor-vine copula with two levels can be derived as\begin{equation*}
\begin{aligned} 
p\left(u_1, \ldots, u_d \mid v_0; \boldsymbol{\theta}\right) & = \prod_{e\{j,k\} \in E_1} p\left(u_{j|v_0}, u_{k|v_0} ; \boldsymbol{\theta_{jk}}\right)  \prod_{i=1}^{d} p\left(u_{i} \mid v_0 ; \boldsymbol{\theta_{0i}}\right) \\
& = \prod_{e\{j,k\} \in E_1} c_{U_{j}, U_k \mid V_0}\left(u_{j \mid v_{0,t} }, u_{k \mid v_{0,t} } ; \boldsymbol{\theta_{jk}} \right) \prod_{i=1}^{d} c_{U_{i}, V_0}\left(u_{i}, v_{0} ; \boldsymbol{\theta_{0i}}\right),\end{aligned}
\end{equation*}
where $E_1$ is the set of edges $e(j, k)$ that connects variable $j$ and variable $k$ at level 2; and $\boldsymbol{\theta} = [\boldsymbol{\theta_{01}^{'}}, \ldots, \boldsymbol{\theta_{0d}^{'}}, \boldsymbol{\theta_{jk}^{'}},\quad \forall e(j, k) \in E_1 ]^{'}$ is the vector of copula parameters.

\newpage
\bibliographystyle{agsm}
\bibliography{WP4}

\end{document}